\documentclass[layout=twocolumn, manuscript=article]{achemso}

\usepackage[version=3]{mhchem} 
\usepackage{lmodern} 
\usepackage{xcolor}
\usepackage{hyperref}
\usepackage{microtype}
\usepackage{lipsum}
\usepackage{amsmath} 
\usepackage{amssymb}
\usepackage{amsfonts}
\usepackage{graphicx}
\usepackage{dcolumn}
\usepackage{bm}
\usepackage[utf8]{inputenc}
\usepackage[T1]{fontenc}
\usepackage{mathptmx}
\usepackage{etoolbox}
\usepackage{dcolumn}
\usepackage{braket}
\def\GG{{\overleftrightarrow{{\bf G}}}}
\usepackage[export]{adjustbox}
\makeatother
\SectionNumbersOn

\title[SSH chain of coreshell nanoparticles]{Coupled electric dipole model for a Su-Schrieffer-Heeger  chain of optically resonant coreshell nanoparticles}
\author{Álvaro Buendía}
 \affiliation{International Iberian Nanotechnology Laboratory (INL), Av. Mte. José Veiga s/n, 4715-330 Braga}
 \email{alvaro.buendia@inl.int}

\author{Nuno M.R. Peres}  \affiliation{International Iberian Nanotechnology Laboratory (INL), Av. Mte. José Veiga s/n, 4715-330 Braga}
\altaffiliation{
Department and Center of Physics  (CF-UM-UP)  of the University of Minho, Campus of Gualtar, 4710-057 Braga.
}
\alsoaffiliation{
POLIMA---Center for Polariton-driven Light--Matter Interactions, University of Southern Denmark, Campusvej 55, DK-5230 Odense M, Denmark.
}%
\email{peres@fisica.uminho.pt}

\begin{document}

\begin{abstract}

Coreshell nanoparticles can combine optical features of different materials in a single nanostructure, which makes them interesting for many applications from biomedicine to energy harvesting. On the other hand, periodic arrays of plasmonic nanoparticles can exhibit topological phenomena such as topological edge states. Here we study periodic chains of Si@Ag coreshell nanoparticles. For this task, we combine the hybridization of surface plasmonic modes in complex nanostructures with the coupled electric dipole formalism employed for modelling the scattering of light by periodic arrays of small plasmonic nanoparticles. We propose treating coreshell nanoparticle as several coupled electric dipoles instead of just one and show how this is a more appropriate framework to study arrays of coreshell nanoparticles, which allows to build a one-to-one connection between the resonant modes of the nanoparticles and the dispersion bands of the system. Within this formalism,  we show that a Su-Schrieffer-Heeger (SSH) chain of coreshell Si@Ag nanoparticles host multiple topological edge states pinned at the resonant frequencies of the coreshell nanoparticles. 

\end{abstract}

\maketitle

\section{Introduction}
The first studies of the scattering of light by small particles were made in the 19th century by Lord Rayleigh. In the first decade of the 20th century, Paul Debye, Gustav Mie and Ludwig Lorenz developed independently a more sofisticated electromagnetic theory, which explained the scattering of light by nanoparticles in terms of electric and magnetic multipoles. 

Metallic nanoparticles have exceptional optical features, produced by the collective vibrations of the electrons in their surface known as localized surface plasmon resonances (LSPR) \cite{Kelly2002}. When they are small and spherical, they scatter approximately light like electric dipoles.

Although plasmonic nanoparticles were originally used for their production of intense colors, more recently they have suscitated interest due to the electric near-field amplification and confinement in subwavelength spaces, which makes them interesting for applications such as enhanced spectroscopy, biosensors and optical cavities \cite{Yu2019}.

Combining  different materials in a single particle such as metals with dielectrics\cite{Sun2020, Barreda2022} or excitonic materials such as organic dyes \cite{Antosiewicz2014, Gentile2017}, makes them even more attractive, allowing them to exploit features from different materials. The applications of coreshell nanoparticles range from biomedicine to energy harvesting. 

Metallic and dielectric nanoparticles have also been used as the building blocks (meta-atoms) of synthetic materials, known as meta-materials, which can present exotical optical phenomena such as negative refractive index. Their two-dimensional version, optical metasurfaces \cite{Meinzer2014, Kuznetsov2024}, have become popular in the last years, due to their ability to control light in unprecedented fashion, substituting bulky devices. Periodic arrays of coreshell nanoparticles could integrate the multifunctional nature of coreshell nanoparticles with the interesting features arising from spatial periodicity, such as topological protection \cite{Ling2015, Rider2019,Rider2022}.

The optical response of plasmonic coreshell nanoparticles or nanomatryushkas has been studied in the frame of hybridization theory \cite{Prodan2003, Ugwuoke2020}. This allows to understand the resonance modes of complex plasmonic nanostructures as the interaction between the LSPRs of metallic spheres and voids. On the other hand, the coupled electric dipole formalism \cite{Purcell1973} have been employed for decades to model the scattering by periodic arrays of metallic and dielectric nanoparticles \cite{Markel1993, Yurkin2007, Rider2019, Abujetas2020, Rider2022}. 

Here we propose to combine the hybridization models of structured nanoparticles with the coupled electric dipole formalism. Instead of considering each coreshell nanoparticle as a single electric dipole, we treat each nanoparticle as two or more coupled-dipoles placed at the same position, which accounts for the hybrdization between the surface modes. We show this is a more appropriate framework, allowing to build a one-to-one connection between resonant modes and dispersion bands, study band properties and predict non-trivial topological phenomena arising from the hybridization in the shells. 

Within this formalism, we show that arranging Si@Ag coreshell nanoparticles in a Su-Schrieffer-Heeger \cite{Su1979} lattice leads to the existence of multiple edge states pinned at two different frequencies which can be tuned by the the filling fraction of coreshell nanoparticle parameters and the host permittivity. This could be employed for non-linear applications and topologically protected generation of higher harmonics.

The manuscript is organized as follows. In Section~\ref{sec:coupled_dipole_polarizability} we propose a coupled electric dipole model for the polarizability of the coreshell nanoparticles. Later in Section~\ref{sec:QS-plasmonic} we study bipartite one-dimensional arrays, which are plasmonic analogues of the infamous Su-Schrieffer-Heeger model, and analyze its topological properties. 
\begin{figure}[ht]
    \centering
        \includegraphics[width=0.99\linewidth]{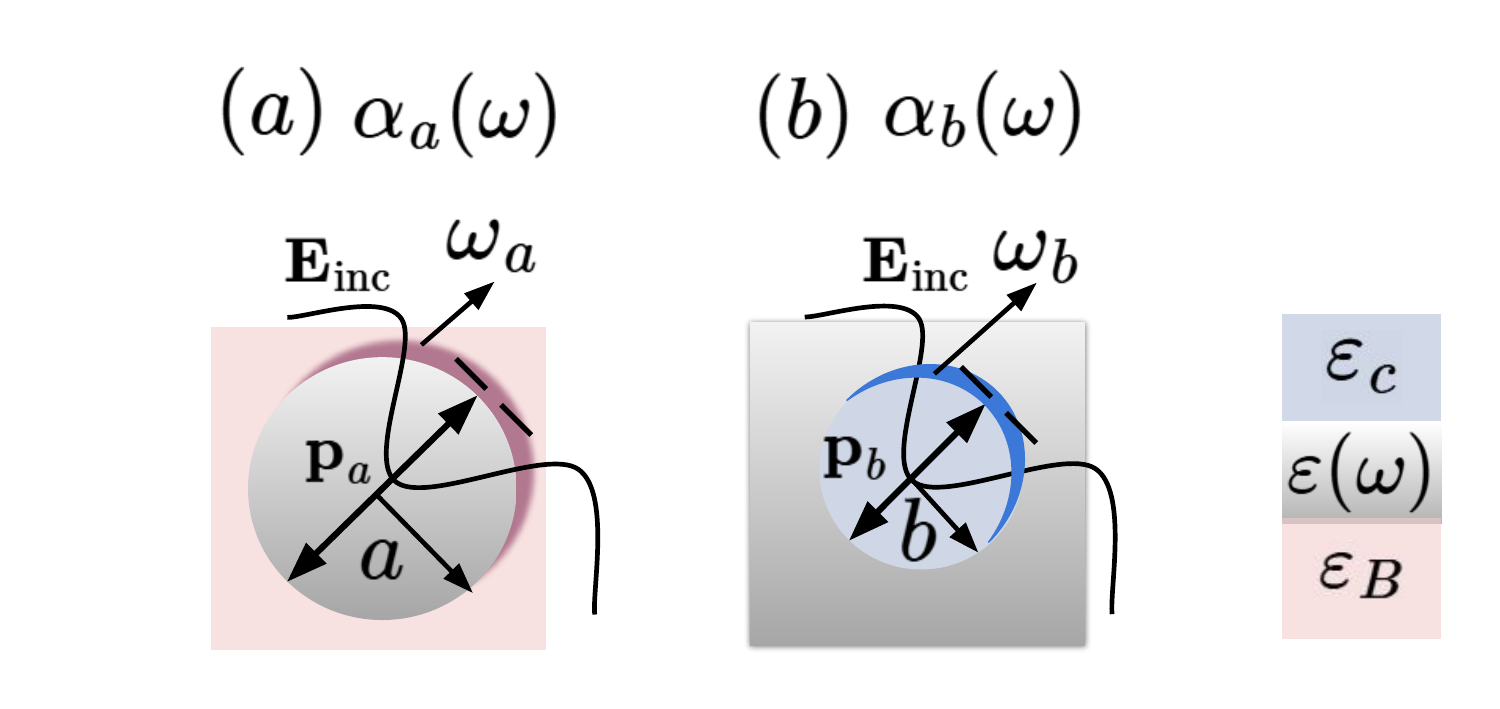}
    \includegraphics[width=0.99\linewidth]{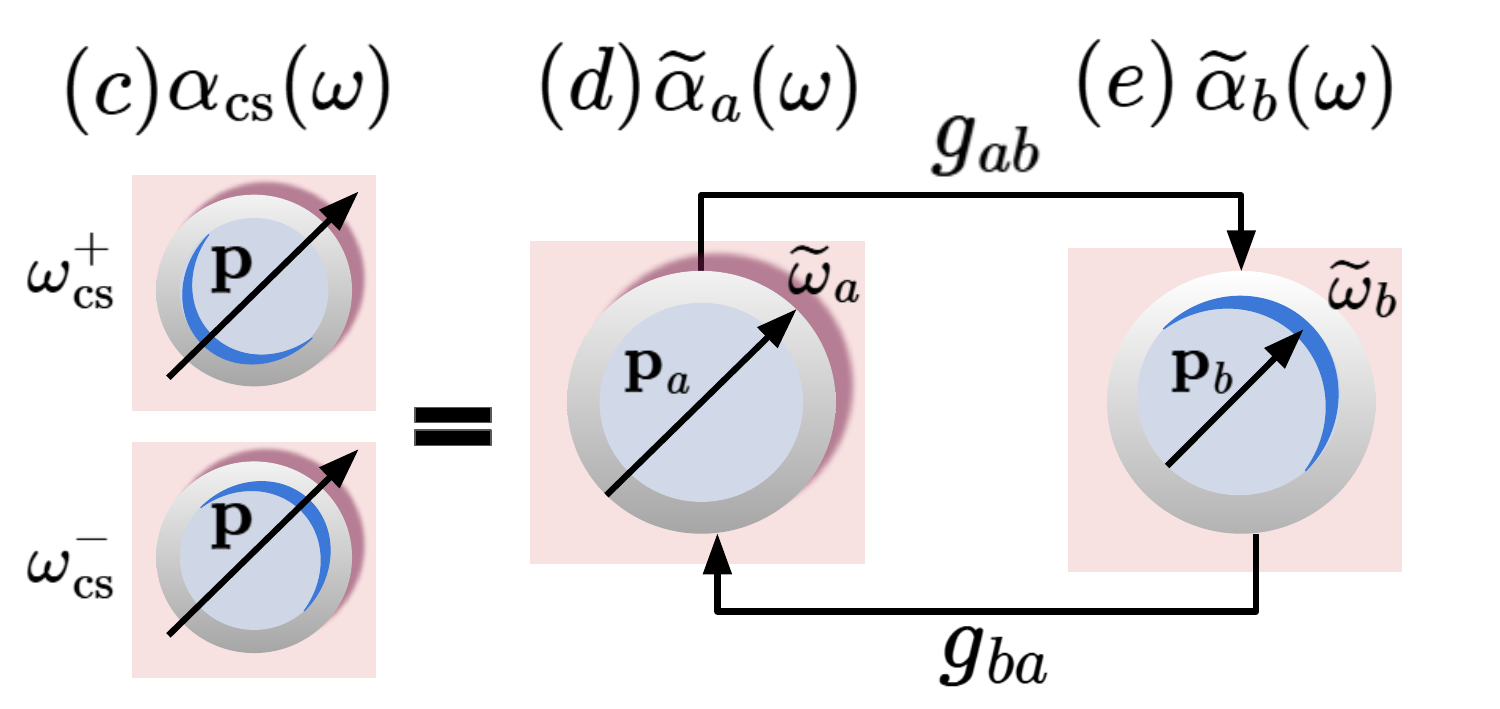}
    \caption{Coreshell polarizability: (a) Localized surface plasmon at a metallic sphere of permittivity $\varepsilon(\omega)$ and radius $a$ in host medium of permittivity $\varepsilon_B$. The external field $E_\textrm{inc}$ induces a dipole $\textbf{p}$. The shaded pink zone represents the oscillating free electrons. (b) Localized surface plasmon at metallic void $\varepsilon(\omega)$ and radius $a$ and permittivity of core $\varepsilon_c$ (c) Localized plasmonic modes at coreshell nanoparticle of radius $a$ and $b$ and core and host permittivity $\varepsilon_B$ and $\varepsilon_c$ (d) and (e) Decomposition of coreshell nanoparticle as the hybridization of the surface sphere and void modes with dipoles $\textbf{p}_a$ and $\textbf{p}_b$.}
    \label{fig:coreshell_optical_response}
\end{figure}
\section{Coupled-dipole model for the polarizability of coreshell nanoparticles}
\label{sec:coupled_dipole_polarizability}
In this section we lay down the concepts needed to understand the SSH-model of coreshell nanoparticles. To fix notation, we briefly review the single particle problem followed by a more detailed analysis of the coreshell particle.
\subsection{Optical resonances of simple spherical nanostructures}
We now study the optical response of spherical nanostructures, starting by a nanosphere embedded in a host medium of permittivity $\varepsilon_B$. 

When the size of the particle is small compared to the wavelength of light $(a < 0.1\lambda)$, a dipolar approximation is accurate and we can neglect higher multipoles. The dipole induced by an incident electric field, $\textbf{E}_\textrm{inc}$, in the nanoparticle is given by\cite{Bohren1998},
\begin{eqnarray}
\textbf{p} = \varepsilon_B \overleftrightarrow{\alpha}(\omega) \textbf{E}_\textrm{inc}, 
\label{eq:dipole_np}
\end{eqnarray}
where $\overleftrightarrow{\alpha}$ is the tensorial electric polarizability of the particle, a $3\times 3$ matrix. For an isotropic sphere, this matrix is proportional to the identity, $\overleftrightarrow{\alpha}(\omega) = \alpha(\omega) \mathbb{I}_3$.

The polarizability of an spherical nanoparticle of permittivity $\varepsilon(\omega)$ of radius $a$ in a host medium of permittivity $\varepsilon_B$ (see Fig.~\ref{fig:coreshell_optical_response}(a)), the dipolar approximation is given by the Clausius-Mosotti relation \cite{Kelly2002, Bohren1998}:
\begin{align}
\alpha_{a}(\omega) = 4\pi a^3 \varepsilon_0 \frac{\varepsilon(\omega) - \varepsilon_\mathrm{B}}{\varepsilon(\omega) + 2\varepsilon_\mathrm{B}}
\end{align}
When the denominator of the polarizability, also known as the Fröhlich function \cite{Ugwuoke2024}, vanish, i.e. $\varepsilon(\omega) = -2\varepsilon_B$, there is an optical resonance. This resonance occurs only on materials with negative permittivity within a frequency window. For metallic nanoparticles it is known as localized surface plasmon resonance (LSPR) and is produced by the collective oscillation of the free electrons in the metallic surface.

If we describe the permittivity of the metal with a Drude model,
\begin{eqnarray}
\varepsilon(\omega) = \varepsilon_\infty  - \frac{\omega_p^2}{\omega^2 +i\omega \gamma}
\end{eqnarray}
where $\omega_{p}$ is the bulk plasmon frequency, $\varepsilon_\infty$ is the permittivity at infinite frequency and $\gamma$ is the optical loss of the material. Then, by solving the Fröhlich condition, ignoring the imaginary part of the permittivity, we find the resonance frequency of the nanosphere,  
\begin{eqnarray}
\omega_a \simeq \frac{\omega_p}{\sqrt{2\varepsilon_B + \varepsilon_\infty}}.
\end{eqnarray}
Now we consider the polarizability of a plasmonic void (see Fig.~\ref{fig:coreshell_optical_response}(b)), where the resonant medium is the medium inside instead of outside. This nanostructure is usually dubbed in literature as cavity, e.g. in \cite{Prodan2003}, however we refrain from using this notation in order to avoid confusion with an optical cavity. We get the polarizability of the void of radius $b$, from the Clausius-Mosotti relation by swapping the permittivity of the two mediums,
\begin{eqnarray}
\alpha_{b}(\omega) = 4\pi b^3 \varepsilon_0 \frac{\varepsilon_c - \varepsilon(\omega)}{\varepsilon_c + 2\varepsilon(\omega)}.
\end{eqnarray}

So the Fröhlich condition in this case is $\varepsilon(\omega) = -\frac{\varepsilon_c}{2}$ and the resonant frequency of the void is 
\begin{eqnarray}
\omega_b \simeq \frac{\omega_p}{\sqrt{\frac{\varepsilon_c}{2} + \varepsilon_\infty}}.
\end{eqnarray}
\subsection{Optical response of coreshell nanoparticles}
Now we consider a nanoparticle with a non-resonant core of radius $b$ with permittivity $\varepsilon_c$ and a resonant shell $\varepsilon(\omega)$ embedded in a medium of constant permittivity $\varepsilon_B$ (see Fig.~\ref{fig:coreshell_optical_response}(c). As initially stated, the resonant modes of complex nanostructures can be understood as the hybridization of modes of simpler structures. In this case, the coreshell modes will be the hybridization between the inner and outer surface modes, i.e. the plasmon modes from the sphere of radius $a$ and the void of radius $b$. (See Figure~\ref{fig:coreshell_optical_response}(a)). This gives rise to two modes, which are the bonding and antibonding modes $\omega_\textrm{cs}^{\pm}$ \cite{Prodan2003}. 

The tensorial polarizability of the coreshell nanoparticle is $\overleftrightarrow{\alpha}_\textrm{cs}(\omega) = \alpha_\textrm{cs}(\omega) \mathbb{I}_3$, where $\alpha_\textrm{cs}(\omega)$ is \cite{Chung2009}
\begin{eqnarray}
&\alpha_{\mathrm{cs}}(\omega) = \nonumber \\ &4\pi\varepsilon_0 a^3 \tfrac{(\varepsilon_c +2\varepsilon(\omega))(\varepsilon(\omega)-\varepsilon_B) + \frac{b^3}{a^3}(\varepsilon_c-\varepsilon(\omega))(2\varepsilon(\omega)+\varepsilon_B)}{(\varepsilon(\omega) + 2\varepsilon_B)(2\varepsilon(\omega) + \varepsilon_c) + 2 \frac{b^3}{a^3}(\varepsilon(\omega)-\varepsilon_B)(\varepsilon_c-\varepsilon(\omega))}.
\label{eq:alphaqs_cs}
\end{eqnarray}
The Fröhlich condition for the coreshell polarizability derives in 
\begin{eqnarray}
\varepsilon(\omega) =  -\varepsilon_{\pm},
\end{eqnarray}
where 
\begin{eqnarray}
 &\varepsilon_\pm = \frac{E_B \mp \sqrt{E_B^2 -4E_AE_C}}{2E_A}, \label{eq:permittivity_coreshell} \\
 &E_A = 2\left(1-\left(\frac{b}{a}\right)^3 \right), \\
 &E_B = 4\varepsilon_B+\varepsilon_c+2\left(\frac{b}{a}\right)^3(\varepsilon_c+\varepsilon_B),\\
 &E_C = 2\varepsilon_c\varepsilon_B \left(1-\left(\frac{b}{a}\right)^3\right).
\end{eqnarray} 
So the resonant frequencies of the coreshell nanoparticle are
\begin{eqnarray}
\omega_\textrm{cs}^{\pm}\simeq \frac{\omega_p}{\sqrt{\varepsilon_\pm + \varepsilon_\infty}}.
\label{eq:omega_cs}
\end{eqnarray} 

The hybrid resonances correspond to the bonding mode $\omega_\textrm{cs}^-$, where the dipoles at the two surfaces are oriented paralel to each other, and the antibonding mode $\Delta\omega_\textrm{cs}^+$, where the dipoles are antiparallel. The frequencies and gap between these modes $\Delta\omega_\textrm{cs} = \omega_\textrm{cs}^+ - \omega_\textrm{cs}^-$ can be tuned through the filling factor $\frac{b}{a}$ and the host permittivity.

Although Equation~\eqref{eq:omega_cs} is specific for Drude materials, the Fröhlich condition for the coreshell nanoparticle is independent of the actual formula of $\varepsilon(\omega)$, so this can be applied not strictly to metals but other type of optically resonant materials, such as excitonic materials like J-aggregates \cite{Gentile2017}.

Now we need the understand the coupling between the inner and outer surface modes. For this, we separate the contributions of the constant and the resonant part of the permittivity, $\varepsilon(\omega) = \varepsilon_\infty + \varepsilon_\textrm{r}(\omega)$. As we see, the void and sphere modes have a homogeneous permittivity for $r>b$ and $r<a$, respectively. In the coreshell nanoparticle, the permittivity is not homogeneous in those two regions of space (see Fig\ref{fig:coreshell_optical_response}(d) and (e)), even far from the resonance frequencies, and there will a trivial shift in the resonance frequencies of the surface modes due to the constant $\varepsilon_\infty$ permittivity. We take in account this shift by considering the quasi-static coreshell polarizability and artificially "turning off" the resonance in one of the surfaces, i.e. substituting $\varepsilon(\omega)$ by $\varepsilon_\infty$ in just the terms arising from one of the surfaces,
\begin{eqnarray}
\widetilde{\alpha}_b(\omega) = 4\pi\varepsilon_0 a^3 \tfrac{(\varepsilon_c +2\varepsilon(\omega))(\varepsilon_\infty-\varepsilon_B) + \frac{b^3}{a^3}(\varepsilon_c-\varepsilon(\omega))(2\varepsilon_\infty+\varepsilon_B)}{(\varepsilon_\infty + 2\varepsilon_B)(2\varepsilon(\omega) + \varepsilon_c) + 2 \frac{b^3}{a^3}(\varepsilon_\infty-\varepsilon_B)(\varepsilon_c-\varepsilon(\omega))}, \\ 
 \widetilde{\alpha}_a(\omega)= 4\pi\varepsilon_0 a^3 \tfrac{(\varepsilon_c +2\varepsilon_\infty)(\varepsilon(\omega)-\varepsilon_B) + \frac{b^3}{a^3}(\varepsilon_c-\varepsilon_\infty)(2\varepsilon(\omega)+\varepsilon_B)}{(\varepsilon(\omega)+ 2\varepsilon_B)(2\varepsilon_\infty + \varepsilon_c) + 2 \frac{b^3}{a^3}(\varepsilon(\omega) - \varepsilon_B)(\varepsilon_c-\varepsilon_\infty)}
\label{eq:alphaqs_cs_shift}.
\end{eqnarray}
The tensorial polarizabilities are just $\overleftrightarrow{\widetilde{\alpha}}_r = \mathbb{I}_3 \widetilde{\alpha}_r(\omega)$. 
The modified resonance frequencies are given by the Fröhlich functions of the two polarizabilities:
\begin{eqnarray}
\varepsilon(\omega) = -\widetilde{\varepsilon}_a = -2\varepsilon_B \frac{2\varepsilon_\infty+\varepsilon_c - \frac{b^3}{a^3}(\varepsilon_c-\varepsilon_\infty)}{2\varepsilon_\infty+ \varepsilon_c + 2\frac{b^3}{a^3}(\varepsilon_c-\varepsilon_\infty)}, \\
\varepsilon(\omega) = -\widetilde{\varepsilon_b}  = -\frac{\varepsilon_c}{2} \frac{2\varepsilon_\infty+\varepsilon_c - 2\frac{b^3}{a^3}(\varepsilon_c-\varepsilon_\infty)}{2\varepsilon_\infty+ \varepsilon_c + \frac{b^3}{a^3}(\varepsilon_c-\varepsilon_\infty)}, 
\end{eqnarray}
\begin{eqnarray}
\widetilde{\omega}_a \simeq \frac{\omega_p}{\sqrt{\widetilde{\varepsilon}_a + \varepsilon_\infty}}, \\
\widetilde{\omega}_b \simeq \frac{\omega_p}{\sqrt{\widetilde{\varepsilon}_b + \varepsilon_\infty}},
\end{eqnarray}
which for $\frac{b}{a}\rightarrow 0$ we see they tend to $\omega_a$ and $\omega_b$, as expected. 

In this description we can define a Rabi splitting between the surface modes. The gap between the coreshell modes $\omega_\textrm{cs}^{\pm}$ and these modified sphere and void modes is entirely produced by the hybridization of the plasmonic modes. When the dephase between the surface modes, $\delta_{ab} = |\widetilde{\omega}_b - \widetilde{\omega}_a| = 0 $, we can define the Rabi splitting,
\begin{eqnarray}
\Omega = \omega_\textrm{cs}^+ - \omega_\textrm{cs}^-.
\end{eqnarray}
For $\frac{b}{a} \rightarrow 0$ the crossing  $\delta_{ab} = 0$ tends to $\varepsilon_B = \frac{\varepsilon_c}{4}$. We see that we can modulate the strength of the coupling by changing $\frac{b}{a}$. For fixed filling fraction $\frac{b}{a}$ and core and shell materials, the gap is only a function of $\varepsilon_B$. 

\begin{figure}[ht]
    \centering
  \includegraphics[width=0.49\linewidth]{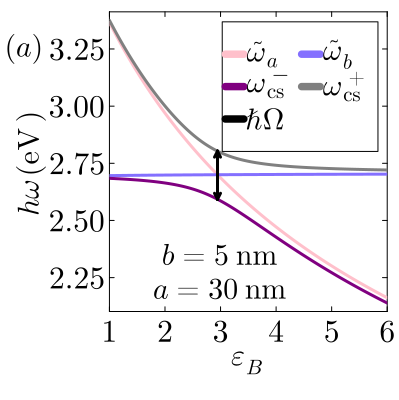}
   \includegraphics[width=0.49\linewidth]{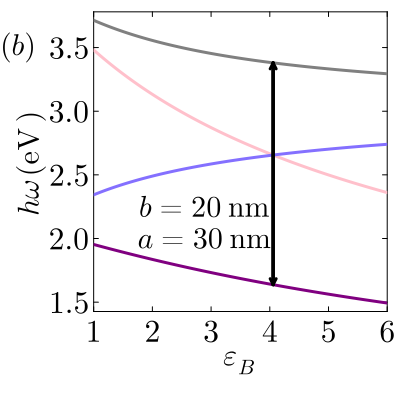}
  \includegraphics[width=0.99\linewidth]{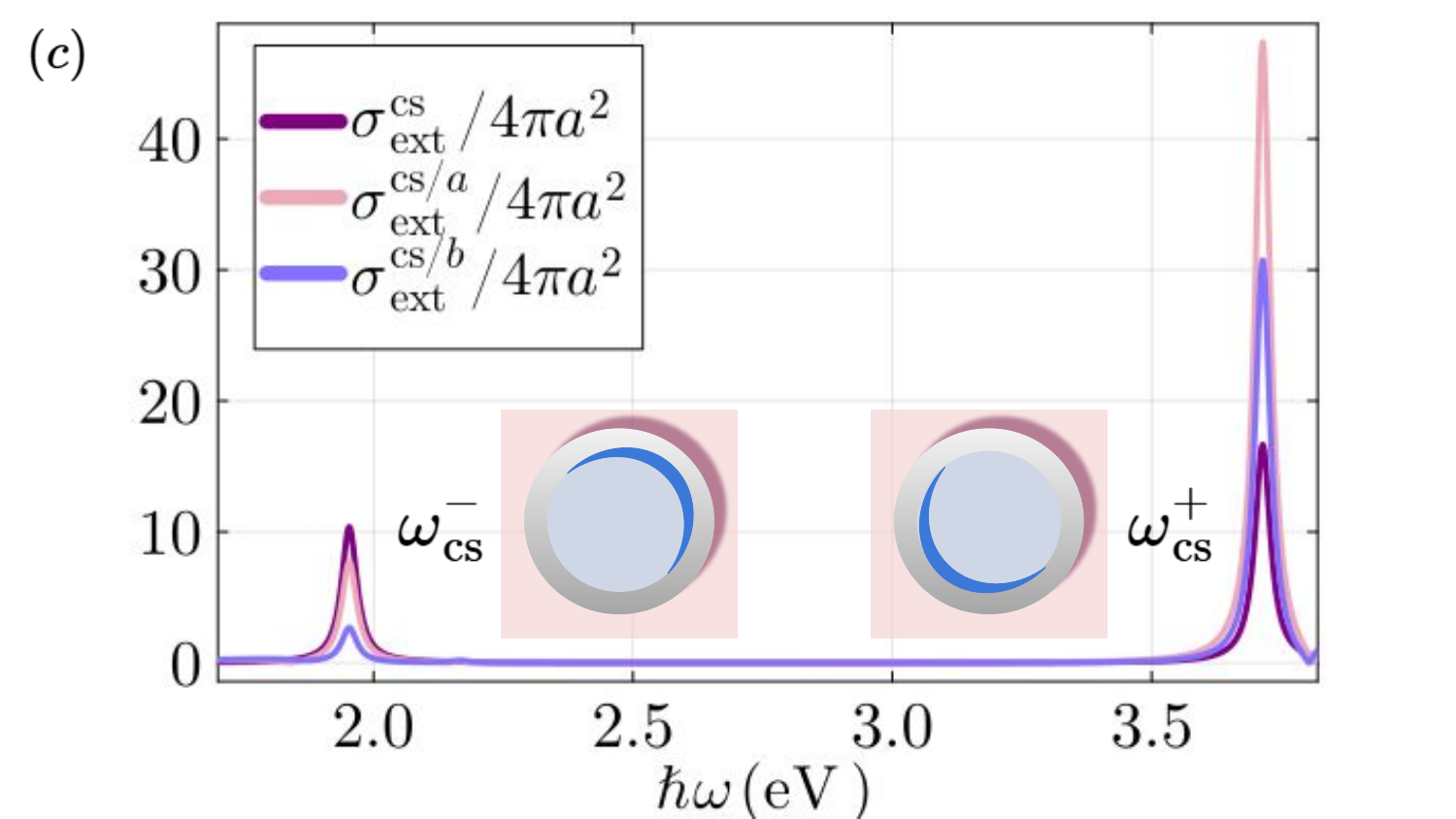}
    \caption{Coupling between surface modes in coreshell nanoparticle: (a) Coupling and Rabi splitting between coreshell modes. The purple and gray lines represent the coreshell modes, while the blue and pink are the bare void and sphere modes $\omega_a$ and $\omega_b$. The black arrows mark the Rabi splitting, $\hbar\Omega = 0.21 $ in this case. We fix $a=30$ nm and $b = 5nm$. Same as (a) but for $b = 20\; \textrm{nm}$. In this case $\hbar\Omega = 1.74$ eV. (c) Optical extinction $\sigma_{ext}(\omega)$ of coreshell nanoparticle : The pink and blue lines represent the optical extinction of a silver sphere on air and Si sphere on silver medium. The purple line is the optical extinction of the coreshell nanoparticle, while the pink and bluea aere the partial optical extinctions at each surface.  For all the plots we used the parameters $\varepsilon_{c} = 11.7, \varepsilon_B = 1, \varepsilon_\infty = 5, \hbar\omega = 8.9~\textrm{eV}, \gamma = 0.0366~\textrm{eV}$ \cite{Vial2005}.}
    \label{fig:coupling_surface_modes}
\end{figure}

In Fig.~\ref{fig:coupling_surface_modes} we study the coupling between the modes depending on $\varepsilon_B$ for two different values of $b$, fixing $a=30$ nm. We set the core material as silicon, $\varepsilon_{c} = 11.7$,  and the shell as silver, $\varepsilon_\infty = 5, \hbar\omega_p = 8.9~\textrm{eV}, \hbar\gamma = 0.0366~\textrm{eV}$ \cite{Vial2005}. In panel (a) we plot the results for $b = 5$ nm. Pink and blue lines are $\widetilde{\omega}_a$ and $\widetilde{\omega}_b$ and the purple and gray lines are $\omega_\textrm{cs}^+$ and $\omega_\textrm{cs}^-$. The black arrows indicate the Rabi splitting. The crossing occurs at $\varepsilon_B = 2.95$ and the Rabi splitting is $\hbar\Omega = 0.21$ eV. We see that far from the crossing, the bands stick to $\widetilde{\omega}_a$ and $\widetilde{\omega}_b$, indicating a small coupling between the plasmonic modes. For $b = 20$ nm, however, plotted in panel (b), the Rabi splitting is $\hbar\Omega = 1.74$ eV at $\varepsilon_B = 4.06$ and the surface modes are strongly hybridized. 

The single dipole description of a coreshell nanoparticle erases information about the coupling between the surface modes. In order to fully grasp the spectral complexity of the particle, and inspired by the previous hybridization model, we attempt to describe the coreshell particle as two coupled dipoles instead of just one. Each dipole represents the electric dipole induced by the surface plasmon at each interface ($r=a$ and $r=b$), $\textbf{p}_a$ and $\textbf{p}_b$ (see Fig. \ref{fig:coreshell_optical_response}(d) and (e)). 

As the shells are concentric, the two dipoles would be placed at the same position, so the total dipole of the nanoparticle would just be $\textbf{p} = \textbf{p}_a + \textbf{p}_b$. In this two-dipole description, each dipole is given by the polarizability multiplied by the incident field plus the scattered field by the other dipole. 
\begin{eqnarray}
&\textbf{p}_a = \varepsilon_B \overleftrightarrow{\widetilde{\alpha}}_a(\omega) (\textbf{E}_\textrm{inc} + \overleftrightarrow{g}_{ba} \textbf{p}_b), \nonumber \\ 
&\textbf{p}_b =  \varepsilon_B\overleftrightarrow{\widetilde{\alpha}}_b(\omega)(\textbf{E}_\textrm{inc} + \overleftrightarrow{g}_{ab} \textbf{p}_a) 
\label{eq:coupled_dipole_coreshell}
\end{eqnarray}

where $g_{ab}$ and $g_{ba}$ are effective couplings between the dipoles, which account for the hybridization between the inner and outer shell modes.

Therefore, in analogy to Eq.~\eqref{eq:dipole_np}  we can express Eq.~\eqref{eq:coupled_dipole_coreshell} as
\begin{eqnarray}
\textbf{P} = \varepsilon_B \overleftrightarrow{A}_\textrm{cs}(\omega) \hat{\textbf{E}}_\textrm{inc},
\label{eq:dipole_cs_np}
\end{eqnarray}
where $\textbf{P} =\begin{pmatrix} \textbf{p}_a \\ \textbf{p}_b \end{pmatrix}$, and $\overleftrightarrow{A}_\textrm{cs}$ is the effective polarizability of the coreshell nanoparticle and $\hat{E}_\textrm{inc}$ is a vector which contains the incident electric field at the position of the sphere and void, i.e. $\mathbf{\hat{\textbf{E}}}_\textrm{inc} = \begin{pmatrix} \textbf{E}_\textrm{inc} \\ \textbf{E}_\textrm{inc} \end{pmatrix}$.
$\overleftrightarrow{A}_\textrm{cs}$ is a polarizability of the coreshell nanoparticle in the coupled-dipole description (see scheme in Fig~\ref{fig:coreshell_optical_response}(d) and (e)). From Equations \eqref{eq:coupled_dipole_coreshell} and \eqref{eq:dipole_cs_np}, and as the couplings must be isotropic, $\overleftrightarrow{g}_{ab/ba} = \mathbb{I}_3 g_{ab/ba}$,
\begin{eqnarray}
\overleftrightarrow{A}_{\textrm{cs}}(\omega) = \mathbb{A}_{\textrm{cs}}(\omega) \otimes \mathbb{I}_3 = \nonumber \\ =\begin{pmatrix} \widetilde{\alpha}_a^{-1}(\omega) & g_{ab} \\ g_{ba} & \widetilde{\alpha}_b^{-1} (\omega)\end{pmatrix}^{-1}  \otimes \mathbb{I}_3 .
\end{eqnarray}
Now, by enforcing the sum of the two induced dipoles in this description is the same as the induced dipole in the single-dipole model,  $p_\mu = p_{a\mu} + p_{b\mu} = \textbf{P}_\mu \cdot\begin{pmatrix} 1 \\ 1 \end{pmatrix}$, where $p_\mu = \textbf{p}\cdot \hat{\boldsymbol{\mu}}$, $p_{r\mu} = \textbf{p}_r \cdot \hat{\boldsymbol{\mu}}$ and {$\textbf{P}_\mu = \begin{pmatrix} p_{a\mu} \\ p_{b\mu} \end{pmatrix}$ and From Eqs. \eqref{eq:dipole_np} and \eqref{eq:dipole_cs_np}, for each polarization $\mu$ we get the same equation $\begin{pmatrix}  1 & 1 \end{pmatrix} \ \mathbb{A}_{\textrm{cs}}(\omega) \begin{pmatrix} 1 \\ 1 \end{pmatrix}  = \alpha_{\textrm{cs}}(\omega)$. From this condition, we find the expression for the effective couplings,
\begin{eqnarray}
 &g_{ab}(\omega) = -g_{ba}(\omega) = \nonumber \\ &\sqrt{\frac{1}{\widetilde{\alpha}_a(\omega)\alpha_\textrm{cs}(\omega)}  + \frac{1}{\widetilde{\alpha}_b(\omega)\alpha_\textrm{cs}(\omega)} - \frac{1}{\widetilde{\alpha}_a(\omega)\widetilde{\alpha}_b(\omega)}}.
\label{eq:effective_couplings}
\end{eqnarray}
If the modes were completely uncoupled, the polarizability of the coreshell nanoparticle would be just the sum of the polarizabilities of the sphere and void, $\alpha_\textrm{cs}(\omega) = \widetilde{\alpha}_a(\omega) + \widetilde{\alpha}_b(\omega)$ and we can see from the last equation that the couplings $g_{ab}, g_{ba}$ would vanish. 


In the two-dipole description we can define partial polarizabilities of the coreshell. For an incident field, $\textbf{E}_\textrm{inc}$, they give the optical response of the sphere or the void modes, separately. These partial polarizabilities can be introduced as projections of the matrix $A_{\textrm{cs}}(\omega)$:
\begin{align}
 \alpha_{\textrm{cs}}(\omega) =  \alpha_{\textrm{cs}}^a(\omega)  +  \alpha_{\textrm{cs}}^b(\omega), \nonumber \\
 \alpha_{\textrm{cs}}^a(\omega) = \begin{pmatrix}  1 & 0 \end{pmatrix} A_{\textrm{cs}}(\omega) \begin{pmatrix} 1 \\ 1 \end{pmatrix}, \\
  \alpha_{\textrm{cs}}^b(\omega) = \begin{pmatrix} 0 & 1 \end{pmatrix} A_{\textrm{cs}}(\omega) \begin{pmatrix} 1 \\ 1 \end{pmatrix}.
\end{align}

In order to study the optical response of the coreshell nanoparticle, we define the extinction cross sections $\sigma_{\textrm{ext}}$\cite{Bohren1998}, which is the sum of absorption $\sigma_{\textrm{abs}}$ and scattering $\sigma_{\textrm{scatt}}$ cross sections, as:
\begin{eqnarray}
\sigma_{\textrm{ext}}^{\textrm{cs}} (\omega) = \frac{k}{\varepsilon_0} \Im(\alpha_{\textrm{cs}}(\omega)) + \frac{k^4}{6\pi\varepsilon_0^2}|\alpha_{\textrm{cs}}(\omega)|^2.
\end{eqnarray}
And we define partial extinction cross sections as: 
\begin{align}
\sigma_{\mathrm{ext}}^{\textrm{cs}/a}(\omega) =  \frac{ k}{\varepsilon_0} \Im(\alpha^a_{\textrm{cs}}(\omega)) + \frac{k^4}{6\pi\varepsilon_0^2}|\alpha^a_{\textrm{cs}}(\omega)|^2 \\
\sigma_{\mathrm{ext}}^{\textrm{cs}/b}(\omega) =  \frac{k}{\varepsilon_0} \Im(\alpha^b_{\textrm{cs}}(\omega)) + \frac{k^4}{6\pi\varepsilon_0^2}|\alpha^b_{\textrm{cs}}(\omega)|^2,
\label{eq:SigmaExt}
\end{align}
In FIG.~\ref{fig:coupling_surface_modes}(c) we plot the extincton of the coreshell nanoparticle $\sigma_\textrm{ext}^\textrm{cs}$ (purple line), and the partial polarizations $\sigma_\textrm{ext}^\textrm{cs/a}$ and $\sigma_\textrm{ext}^\textrm{cs/b}$ (pink and blue lines).  We set air as the host medium, $\varepsilon_B = 1$, for which the resonance frequencies of the coreshell nanoparticle are $\hbar\omega_{\textrm{cs}}^{-} =  1.95~$eV and $\hbar\omega_{\textrm{cs}}^{+} = 3.71~$eV.  Comparing the partial and the total extinction of the nanoparticle, we can see for the lower frequency mode $\omega_\textrm{cs}^-$, there is a constructive interference between the cavity and sphere modes, indicating it is the bonding mode, while for the upper mode $\omega_\textrm{cs}^+$, there is destructive interference, so this is an antibonding mode. We also see the weights are not equal for the cavity and void, as there is a dephase between the modes and the intensity of the sphere mode is proportional to $a$ while for the void is proportional to $b$. 

This multi-dipole decomposition of coreshell nanoparticles could be generalized for nanostructures with more resonances, like nanomatryushkas or coreshell nanoparticles with resonant core and shell \cite{Antosiewicz2014}, as we address on Appendix \ref{sec:coreshell_decomp}.
\section{Quasi-static plasmonic SSH chain of coreshell nanoparticles}
\label{sec:QS-plasmonic}
\begin{figure}[!h]
    \centering
    \includegraphics[width=0.99\linewidth]{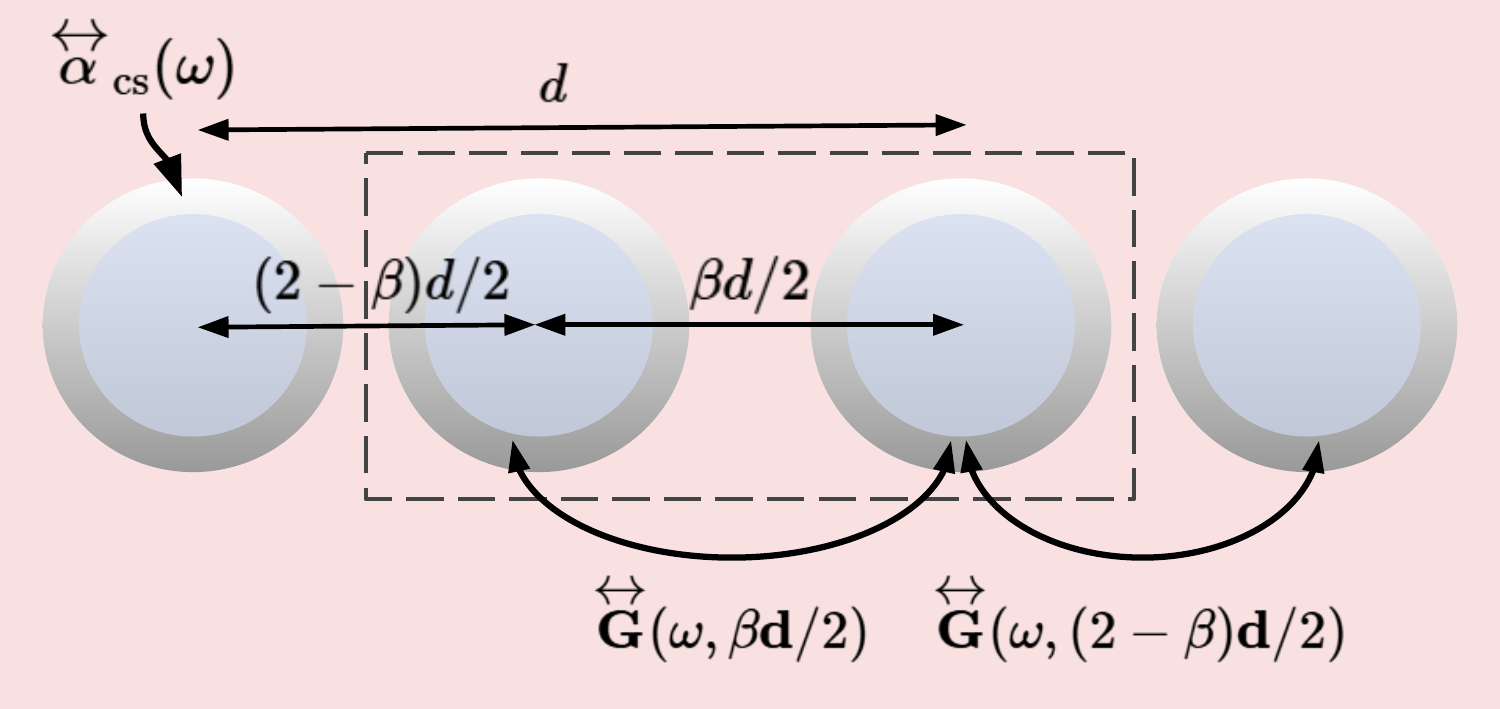}
    \includegraphics[width=0.99\linewidth]{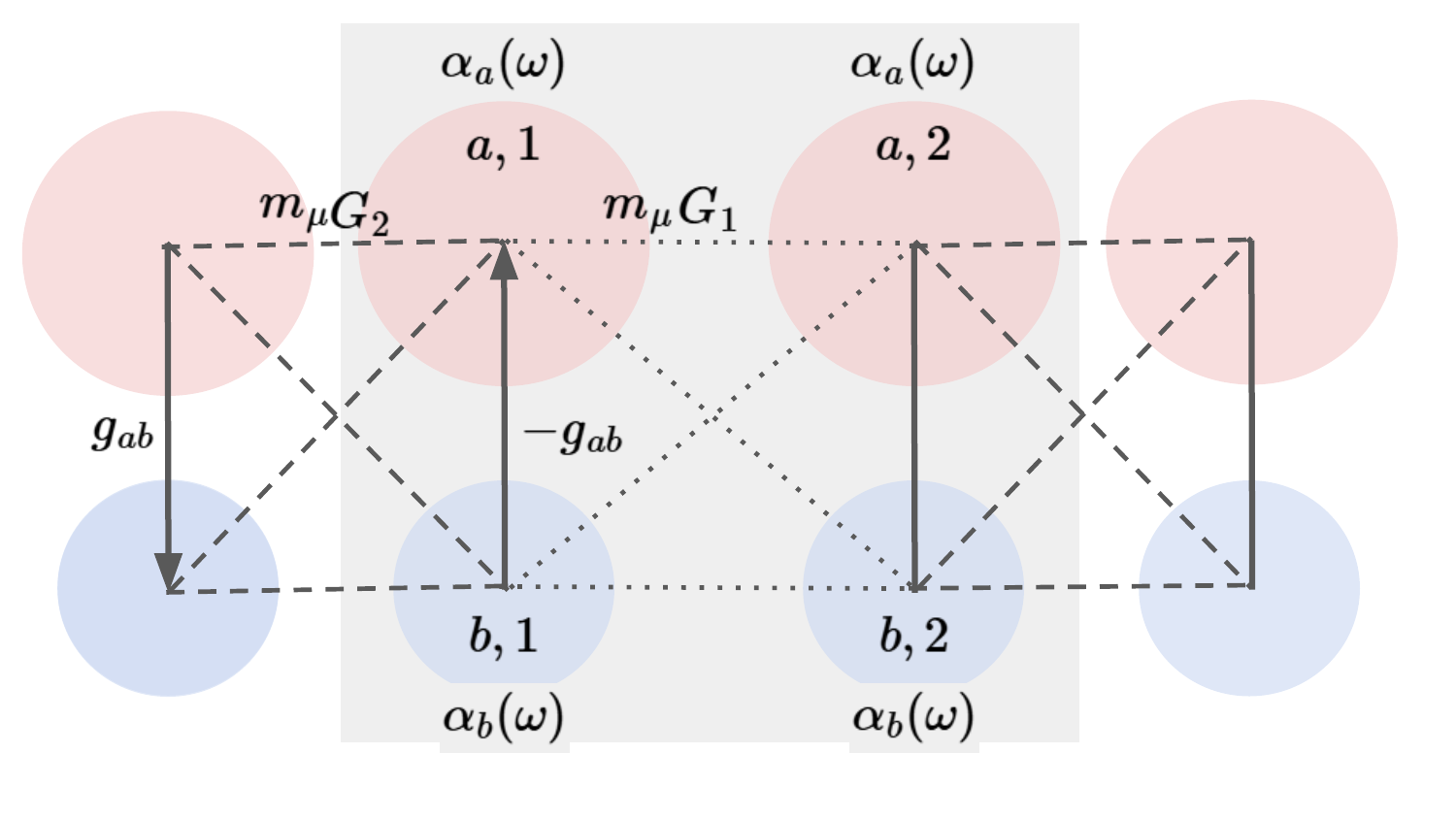}
    \caption{SSH chain of coreshell nanoparticles: The SSH chain is a periodic 1D array with two particles per unit cell, with staggered couplings due to alternating distances. The unit cell is delimited by the dashed gray line. The distance between the particles in the unit cell is $\beta d^2$ where $\beta$ is a dimensionless parameter between 0 and 2, while the distance between first neighbours from adjoint cells is $(2-\beta) d/¨2$. The couplings between the particles are mediated by the Green's dipole-dipole function $\overleftrightarrow{\textbf{G}}(\omega,\textbf{R})$. (b) Mapping of the coreshell nanoparticle chain to a bilayer SSH chain. The solid, dashed and dotted lines represent the intraparticle coupling $g_{ab}$, the intracell coupling, $G_1$ and the intercell coupling, $G_2$. }
    \label{fig:SSH_cs}
\end{figure}
The Su-Schrieffer-Heeger (SSH) model was initially introduced to explain the physics of polyacetylene chains \cite{Su1979}. It is a tight binding model with two particles per unit cell and staggered hoppings. Later it was used as a toy model for 1D topological insulators \cite{Asboth2016}. When arranged in a finite array commensured with the appropriate unit cell, the SSH chain host topological 0D edge states, also dubbed end states, which are exponentially localized at the boundaries of the chains. These edge states are protected against disorder or perturbations by its symmetries.

Due to its simplicity, in the last decades there have been proposals of analogues of the SSH model in different platforms, such as mechanical, acoustic \cite{Coutant2021}, and optical systems \cite{Orsini2024}.

In the quasi-static limit, a bipartite lattice of plasmonic nanoparticle mimics the SSH tight binding model, as it was firstly proposed in References \cite{Ling2015} and \cite{Downing2017}. However, further studies showed that the long-range interactions in transversal polarization can affect the topological properties of the system and can spoil the protection of the edge states \cite{Pocock2018, Pocock2019}. Later, there have been other theoretical proposals and experimental realizations of the SSH model in plasmonics beyond nanospheres, such as chains of nanodisks \cite{Poddubny2014}, split rings \cite{Shao2025}, nanoslits \cite{Schurr2025} and nanospheroids \cite{Buendia2023}. Other proposed devices involving graphene plasmons, such as stacked graphene nanoribbons \cite{Xia2023} or plasmonic crystal made of metallic rods on top of a graphene sheet \cite{Rappoport2021}.

In this study we restrict to a quasi-static, first neighbour approximation for a chain of plasmonic coreshell nanospheres. However, in Appendix~\ref{sec:long-range}, we show how to make the retardation and radiation corrections to the polarizability and Green's dipole-dipole functions. For more information on this regime, the reader can check References \cite{Pocock2018,Pocock2019} . 

The plasmonic SSH array is a chain of metallic nanospheres with two alternate distances: the intracell distance $\beta\frac{d}{2}$ and the intercell distance $\left(2-\beta\right)\frac{d}{2}$, $d$ being the size of the unit cell (see scheme in Figure~\ref{fig:SSH_cs}).

For an array of dipoles,  each dipole is determined by the incident electric field plus the sum of the scattered fields by the other particles. The  field radiated by a particle in the dipolar approximation is given by \cite{Bohren1998}
\begin{eqnarray}
\textbf{E} =  \frac{k^2}{\varepsilon_0\varepsilon_B} \overleftrightarrow{\textbf{G}}(\omega,\textbf{R})\textbf{p}, 
\end{eqnarray}
where $\overleftrightarrow{\textbf{G}}(\omega,\textbf{R})$ is the Green's dyadic  dipole-dipole function: 
\begin{eqnarray}
\GG(\omega, \mathbf{R}) = \frac{e^{ikR}}{4\pi R}\bigg[\left(1 + \frac{i}{kR} - \frac{1}{k^2 R^2}\right)\mathbb{I}_3 - \\ \nonumber \left(1 + \frac{3i}{kR} -\frac{3}{k^2 R^2}\right)\frac{\textbf{R}\otimes \textbf{R}}{R^2} \bigg].
\label{eq:dyadic_G}
\end{eqnarray}
Therefore, considering each dipole $\textbf{p}_n$ is ar position $\textbf{r}_n$, in the absence of incident electric field, we have the self-consistent coupled-dipole equations,
\begin{eqnarray}
\overleftrightarrow{\alpha}^{-1}_n(\omega)\textbf{p}_n =  \frac{k^2}{\varepsilon_0} \sum_{m \neq n} \overleftrightarrow{\textbf{G}}(\omega,\textbf{R}
_{nm}) \cdot \textbf{p}_m.
\label{eq:eqGiso}
\end{eqnarray}
where $\textbf{R}_{nm} = \textbf{r}_n -\textbf{r}_m$.

The coupled-dipole approximation is accurate as long as the gap between nanoparticles is at least of the order of the radius of the particles, or the distance between particles is $R>3a$. For distances smaller than this, the contribution by higher multipoles to the interaction between two particles should cannot be neglected. 

In the quasi-static limit, when $kR \ll 1$, the short-range term $\propto R^{-3}$ dominates and the Green's function can be approximated by: 
\begin{eqnarray} 
\GG_{\mathrm{QS}}(\omega, \textbf{R}) = \frac{1}{4\pi k^2 R^3}\left[-\mathbb{I}_3 +3\frac{\textbf{R}\otimes \textbf{R}}{R^2}  \right] = \nonumber \\ =  G_\textrm{QS}(R) \mathbb{M}_\textrm{pol}(\textbf{R}). 
\label{eq:dyadic_G_QS}
\end{eqnarray}
where we define $G_\textrm{QS}(R) =  \frac{1}{4\pi k^2 R^3}$ and $\mathbb{M}_\textrm{pol}(\textbf{R})= \left[-\mathbb{I}_3 +3\frac{\textbf{R}\otimes \textbf{R}}{R^2}  \right]$. 

Now we imagine a crystalline finite array of coreshell nanoparticles with $n_c$ unit cells and $\mathcal{N}$ particles per unit cell. Analogously to Eq. \eqref{eq:eqGiso}, for the two-dipole description of the polarizability of a single coreshell nanoparticle, the coupled-dipole self-consistent equations are now: 
\begin{eqnarray}
\overleftrightarrow{\widetilde{\alpha}}_r^{-1}(\omega) \textbf{p}_{r,s,n} = \sum_{r\neq r'} \overleftrightarrow{g}_{rr'}\textbf{p}_{r',s,n} \nonumber \\ + \frac{k^2}{\varepsilon_0}  \sum\limits_{ r',s',m}    ^*\overleftrightarrow{\textbf{G}}(\omega,\textbf{R}_{nm,ss'}) \textbf{p}_{r',s,m} \;
\end{eqnarray}
where $r = {a,b}$ and $\textbf{R}_{nm,ss'}$ is the relative position between the particle in unit cell $n$ and sublattice $s$ and the particle in unit cell $m$ and sublattice $s'$. The first term involves the couplings between the dipoles in a single particle, $\mathbb{I}_3 g_{ab}$ and $\mathbb{I}_3 g_{ba}$ from Equation~\eqref{eq:effective_couplings}, while the second term accounts for the couplings between the dipoles of different particles and the $*$ means it excludes the self-interaction term $n=m$ and $s=s'$. 

For the SSH lattice, in the quasi-static regime we can also take a first neighbours approximation, due to the interaction decaying rapidly in space. The coupled dipole equations in \eqref{eq:eqGiso} are reduced to
\begin{eqnarray}
&\overleftrightarrow{\widetilde{\alpha}}_a^{-1}(\omega) \textbf{p}_{a,1,n} = -\mathbb{I}_3 g_{ab}\textbf{p}_{b,1,n} + \nonumber \\ &\mathbb{M}_\textrm{pol} \sum\limits_{r'} G_1 \textbf{p}_{r',2,n} + \theta(n-2) G_2 \textbf{p}_{r',2.n-1} \nonumber \\ &\overleftrightarrow{\widetilde{\alpha}}_b^{-1}(\omega) \textbf{p}_{b,1,n} = \mathbb{I}_3 g_{ab}\textbf{p}_{a,1,n} + \nonumber \\ &\mathbb{M}_\textrm{pol}    \sum\limits_{r'} G_1 \textbf{p}_{r',2,n} + \theta(n-2) G_2 \textbf{p}_{r',2.n-1} \nonumber \\ &\overleftrightarrow{\widetilde{\alpha}}_a^{-1}(\omega) \textbf{p}_{a,2,n} = -\mathbb{I}_3 g_{ab}\textbf{p}_{b,2,n} +\nonumber \\  &\mathbb{M}_\textrm{pol}    \sum\limits_{r'} G_1 \textbf{p}_{r',2,n} + \theta(nc-n-1) G_2 \textbf{p}_{r',2,n+1} \nonumber \\ &\overleftrightarrow{\widetilde{\alpha}}_b^{-1}(\omega) \textbf{p}_{b,2,n} = \mathbb{I}_3 g_{ab}\textbf{p}_{a,2,n} + \nonumber \\ &\mathbb{M}_\textrm{pol}    \sum\limits_{r'} G_1 \textbf{p}_{r',2,n} + \theta(n_c-n-1) G_2 \textbf{p}_{r',2.n+1} \nonumber
\label{eq:system_SSH_equations}
\end{eqnarray}
where $\theta(x)$ is the Heaviside step function and $G_{1,2}$ are the Green dipole-dipole interactions (normalized by $\frac{\varepsilon_0}{k^2}$) for the two inter-particle distances, $\beta d/2$ and $(2-\beta)d/2$, this is $G_1 = \frac{k^2}{\varepsilon_0} G_\textrm{QS}(\beta d/2) = \frac{1}{4\pi\varepsilon_0(\beta d/2)^3}$,  $G_2 = \frac{k^2}{\varepsilon_0} G_\textrm{QS}((2-\beta )d/2 ) = \frac{1}{4\pi \varepsilon_0 ((2-\beta) d/2)^3}$. We see at the first and last cells, the $G_2$ term nulls, due to the translation symmetry breaking. 

These equations can be reformulated in a matricial way as \cite{Rider2022}:
\begin{eqnarray}
\mathbb{A}^{-1}(\omega) \mathbf{P}_{f} = \mathbb{G} \mathbf{P}_{f}.
\label{eq:eigensystem_finite_SSH}
\end{eqnarray}

where $\mathbb{G}$ and $\mathbb{A}$ are $3n_\textrm{res}n_c \mathcal{N} \times 3n_\textrm{res}n_c \mathcal{N}$ matrices, where $n_\textrm{res}$ is the number of resonances/dipoles per single particle (in this case $n_\textrm{res} = 2$). The Green matrix is
\begin{eqnarray}
\mathbb{G} = \begin{pmatrix} G_{xx} & G_{xy} & G_{xz} \\ G_{yx} & G_{yy} & G_{yz} \\ G_{zx} & G_{zy} & G_{zz} \end{pmatrix} =\mathbb{G}_0 \otimes \mathbb{M}_\textrm{res} \otimes \mathbb{M}_\textrm{pol}(\hat{\bf{x}})    = \nonumber \\ \mathbb{G}_0  \otimes \begin{pmatrix} 1& 1 \\ 1 & 1\end{pmatrix}  \otimes \begin{pmatrix} 2 & 0 & 0 \\ 0 & -1 & 0 \\ 0 & 0 & -1 \end{pmatrix}
\label{eq:Gmatrix_fin}
\end{eqnarray}
being $\mathbb{G}_{0}$ a $n_c\mathcal{N}\times n_c\mathcal{N}$ matrix. 
The general expression for $\mathbb{G}_0$ is:
\begin{eqnarray}
\mathbb{G}_0 = \mathbb{I}_{n_c}\otimes \begin{pmatrix} 0 & G_1 \\ G_1 & 0 \end{pmatrix} \nonumber \\ + 0 \oplus \left(\mathbb{I}_{n_c-1} \otimes \begin{pmatrix} 0 & G_2 \\ G_2 & 0 \end{pmatrix} \right) \oplus 0 
\end{eqnarray}
So for $n_c=2$, like in Figure \ref{fig:SSH_cs}(a), $\mathbb{G}_0$ is
\begin{eqnarray}
\mathbb{G}_{0} = \begin{pmatrix} 0 & G_{1} & 0 & 0  \\ G_1 & 0 & G_2 & 0 \\ 0 & G_2 & 0 & G_1  \\ 0 & 0 & G_1  & 0 \end{pmatrix},
\end{eqnarray}
%
The polarizability matrix is:
\begin{eqnarray}
\mathbb{A}^{-1}(\omega) = 
\mathbb{I}_{n_c} \otimes \mathbb{I}_2  \otimes \mathbb{A}^{-1}_\textrm{cs}(\omega)\otimes \mathbb{I}_3 
\end{eqnarray}
Due to the 1D nature of the array and the spherical shape of the particles, the dipoles of different particles which are oriented in perpendicular polarizations do not interact and this leads to the Green matrix of the finite array (Equation~\eqref{eq:Gmatrix_fin}) and the matrix $\mathbb{A}$ being block-diagonal. 

Finally, the vector $\textbf{P}_f$ contains the dipolar momentums of all the dipoles,
\begin{eqnarray}
\textbf{P}_f =  \begin{pmatrix} \textbf{p}_{a,1,1} \\ \textbf{p}_{b,1,1} \\ \textbf{p}_{a,2,1} \\ \textbf{p}_{b,2,1} \\ \vdots \\ \textbf{p}_{a,2,n_c} \\   \textbf{p}_{b,2,n_c} \end{pmatrix},
\end{eqnarray}
and they can be obtained solving Equation~\eqref{eq:eigensystem_finite_SSH}, i.e. finding the eigenmodes of $\mathbb{A}^{-1}(\omega) - \mathbb{G}$.  

Now we analyze the chain with periodic boundary conditions. For an infinite periodic array, the dipolar momentums of the particles will respect the Bloch periodicity, this is $\textbf{p}_{r,n+2m} = \textbf{p}_{r,n}e^{-iqmd}$. By plugging in this into the coupled dipole equations we have, 
\begin{align}
&\overleftrightarrow{\widetilde{\alpha}}_r^{-1}(\omega) \textbf{p}_{r,s} = \sum_{r\neq r'} \overleftrightarrow{g}_{rr'}\textbf{p}_{r',s} \nonumber \\ &+ \frac{k^2}{\varepsilon_0}  \sum\limits_{ m\neq 0, r'}\overleftrightarrow{\textbf{G}}(\omega,md ) \textbf{p}_{r',s}e^{iqmd}   \nonumber\\ &+\frac{k^2}{\varepsilon_0}  \sum\limits_{m,r,s'\neq s}\overleftrightarrow{\textbf{G}}(\omega, md  +(-1)^{s'} \beta d/2 ) \textbf{p}_{r',s'}e^{iqmd}.
\label{eq:coupled_dipole_cs}
\end{align}

So now we see that we have only 4 independent dipoles, accounting for the two particle per unit cell and the two resonances per particle.  

In the quasi-static, nearest-neighbours approximation, the coupled dipole equations are just
\begin{eqnarray}
\overleftrightarrow{\widetilde{\alpha}}_a^{-1}(\omega) \textbf{p}_{a,1} = -\mathbb{I}_3 g_{ab}\textbf{p}_{b,1} \nonumber \\ + \mathbb{M}_\textrm{pol}    \sum\limits_{r'}\left(G_1 + G_2 e^{iqd}\right)\textbf{p}_{r',2}  \nonumber \\ \overleftrightarrow{\widetilde{\alpha}}_a^{-1}(\omega) \textbf{p}_{a,2} = -\mathbb{I}_3 g_{ab}\textbf{p}_{b,2} \nonumber \\ + \mathbb{M}_\textrm{pol}    \sum\limits_{r'}\left(G_1 + G_2 e^{-iqd}\right)\textbf{p}_{r',1} \nonumber \\ 
\overleftrightarrow{\widetilde{\alpha}}_b^{-1}(\omega) \textbf{p}_{b,1} = \mathbb{I}_3 g_{ab}\textbf{p}_{a,1} \nonumber \\ + \mathbb{M}_\textrm{pol}    \sum\limits_{r'}\left(G_1 + G_2 e^{iqd}\right)\textbf{p}_{r',2} \nonumber \\
\overleftrightarrow{\widetilde{\alpha}}_b^{-1}(\omega) \textbf{p}_{b,2} = \mathbb{I}_3 g_{ab}\textbf{p}_{a,1} \nonumber \\ + \mathbb{M}_\textrm{pol}    \sum\limits_{r'}\left(G_1 + G_2 e^{-iqd}\right)\textbf{p}_{r',1}
\end{eqnarray}

We can condense the coupled equations in a matricial way as, 
\begin{eqnarray}
\mathcal{G}(q) \mathcal{P}(q)= \mathcal{A}^{-1}(\omega) \mathcal{P}(q),
\label{eq:CDBloch}
\end{eqnarray}
where $\mathcal{G}$ and $\mathcal{A}$ are the Green and polarizability matrices, in this case they are of dimension $3n_{res}\mathcal{N} \times 3n_{res}\mathcal{N}$. We now define the Green's matrix and polarizability of the periodic array,
\begin{eqnarray}
&\mathcal{G}(q) =   \mathcal{G}_0(q) \otimes \mathbb{M}_\textrm{res} \otimes  \mathbb{M}_{\textrm{pol}}(\hat{x}) = \nonumber \\ = &\begin{pmatrix} 0 & g(q)\\ g^\dag(q) & 0 \end{pmatrix}  \otimes \begin{pmatrix} 1 & 1 \\ 1 & 1 \end{pmatrix} \otimes \begin{pmatrix} 2 & 0 & 0 \\ 0 & -1 & 0 \\ 0 & 0 & -1 \end{pmatrix}, \label{eq:G_matrix_periodic} \end{eqnarray}
where $g(q) = G_1 + G_2 e^{-iqd}$. The polarizability matrix is
\begin{eqnarray}
 \mathcal{A}^{-1}(\omega) =   \mathbb{I}_2  \otimes \mathbb{A}_\textrm{cs}^{-1}(\omega)\otimes \mathbb{I}_3,
\end{eqnarray}
Explicitly,
\begin{eqnarray}
\mathcal{P}(q) =  \begin{pmatrix} \textbf{p}_{a,1}(q) \\ \textbf{p}_{b,1}(q) \\ \textbf{p}_{a,2}(q) \\ \textbf{p}_{b,2}(q)
\end{pmatrix}. \label{eq:eigen_infinite}
\end{eqnarray}

As the polarization modes are uncoupled, we can treat each polarization separately. For each polarization $\mu$ we have

\begin{eqnarray}
&\mathcal{A}^{-1}_\mu(\omega) - \mathcal{G}_\mu(q) = \nonumber \\ &\begin{pmatrix} \widetilde{\alpha}_a^{-1}(\omega) & g_{ab} & g_{\mu}(q) & g_{\mu}(q) \\ -g_{ab} & \widetilde{\alpha}_b^{-1}(\omega) & g_\mu(q) & g_\mu(q) \\ g^\dag_\mu(q) & g^\dag_\mu(q) & \widetilde{\alpha}_a^{-1}(\omega) & g_{ab} \\ g^\dag_\mu(q) & g^\dag_\mu(q) & -g_{ab} & \widetilde{\alpha}_b^{-1}(\omega) \end{pmatrix}
\label{eq:coupled_ssh_mapping}
\end{eqnarray}

where $A_\mu$ and $G_\mu$ are the polarization projections of the polarizability and Green's matrix and $g_\mu = m_\mu g(q)$ where $m_\mu  = [\mathbb{M}_{\textrm{pol}}(\hat{x})]_{\mu\mu}$, i.e. $m_x = 2$ for the longitudinal mode, and $m_{y,z} = -1$ for the two transversal modes.

In Reference \cite{Chen2023} a photonic crystal of I-shaped slabs arranged on a SSH lattice is mapped to a bilayer SSH chain, also known as two-leg ladder SSH. Analogously we can make a similar mapping in our system. For each polarization, and as we represent in Fig.~\ref{fig:SSH_cs}(b) the matrix in Equation~\ref{eq:coupled_ssh_mapping} can be mapped to a tight binding Hamiltonian of two coupled SSH chains, where the interlayer interactions are $g_{ab}$ for the same sublattice and the intracell and intercell couplings are $G_1$ and $G_2$ for the same or different layers and on-site energies are $\widetilde{\alpha}^{-1}_{a,b}(\omega)$ for $a$ and $b$ sites. 

This shines light of how monolayer arrays of structured or complex-shape nanoparticles could be mapped to multilayer tight binding lattices, whose topology have been previously studied \cite{Padavic2018, Guo2025, Chen2023}.

One of the characteristics of multilayer SSH models is the presence of edge states at different gaps, which as we will show in Section \ref{subsec:spec} are also present in this system. The number of effective layers in the array could be increased just by adding resonant layers in the nanoparticle (either the core or extra shells). 

As we have emphasized previously, for a SSH chain of spherical nanoparticles, polarization modes are uncoupled. However, breaking the 1D character of the chain (as in zigzag lattices) or the axial symmetries of the particles leads to the hybridization of polarization modes and more effective coupled layers in the analogous tight binding model.

\subsection{Dispersion bands}
The dispersion bands can be derived by solving equation \eqref{eq:CDBloch}, or alternatively,
\begin{eqnarray}
\det\left(\mathcal{A}^{-1}(\omega) - \mathcal{G}(q)\right) = 0. 
\end{eqnarray}
As the determinant is the product of eigenvalues, this equation can be decomposed as a system of equations, 
\begin{eqnarray}
\lambda_i \left(\mathcal{A}^{-1}(\omega) - \mathcal{G}(q)\right) = 0,
\end{eqnarray}
where $\lambda_i(M)$ is the $i$-th eigenvalue of matrix $M$. Instead of solving a single equation with many solutions, we can now calculate each band separately. As we decomposed each coreshell nanoparticle as two dipoles, there is now a one-to-one correspondence between the number of equations and the number of bands, which is the utility of this description. The number of bands matches with the number of degrees of freedom. In this case, we have 3 polarizations, 2 resonances per particle and 2 lattice sites, so we will have $ 3n_\textrm{res}\mathcal{N} = 3\cdot 2\cdot 2 = 12$ modes. As the lattice is $1D$ and the particles are isotropic, all the transversal bands will be doubly degenerate. 

The solutions of these equations can be found numerically, with a non-linear solver. However, in order to get analytical expressions for the bands, we can use the single-dipole description for coreshell nanoparticles. In this model, the dispersion bands are just the solutions of $\det(\alpha_\textrm{cs}^{-1}(\omega)\mathbb{I}_2 - m_\mu \mathcal{G}_0(q))$, i.e. $\alpha^{-1}_\textrm{cs}(\omega) = \pm |g_\mu(q)|$,where  This derives in the condition $\varepsilon(\omega) = -\varepsilon_{\pm,\mu, n}(q)$, where
\begin{eqnarray}
 &\varepsilon_{\mu,\pm,n}(q) = \frac{\widetilde{E}_B \mp \sqrt{\widetilde{E}_B^2 -4\widetilde{E}_C\widetilde{E}_A}}{2\widetilde{E}_A}, \label{eq:permittivity_dispersion} \\
 &\widetilde{E}_A = E_A\left(1 +(-1)^{n+1} 4\pi a^3\varepsilon_0 |g_\mu(q)||\right), \\
 &\widetilde{E}_B = E_B\left(1 +(-1)^{n+1} 4\pi a^3\varepsilon_0|g_\mu(q)|\right)+ \nonumber \\ &(-1)^n 12\pi a^3 \varepsilon_0\varepsilon_B |g_\mu(q)|\left(2+\left(\frac{b}{a}\right)^3\right),\\
 &\widetilde{E}_C = E_C\left(1 +(-1)^{n} 2\pi a^3\varepsilon_0|g_\mu(q)|\right).
 \end{eqnarray}

The dispersion bands are then just
\begin{eqnarray}
\omega_{\mu,\pm,n}(q) = &\frac{\omega_p}{\sqrt{\varepsilon_{\mu,\pm,n}(q) + \varepsilon_\infty}} = \nonumber \\ = &\omega_{cs}^{ \pm}\sqrt{\frac{\varepsilon_{\pm} + \varepsilon_\infty}{\varepsilon_{\mu,\pm,n}(q) + \varepsilon_\infty}}.
\label{eq:dispersion_bands}
\end{eqnarray} 
We can see that in the limit of isolated nanoparticles, when $d\gg a$, $4\pi\varepsilon_0 a^3 |g_\mu(q)| \ll 1$, so Eq.~\eqref{eq:permittivity_dispersion} tends to Eq.~\eqref{eq:permittivity_coreshell} and the dispersion bands tend to the frequencies of the isolated coreshell nanoparticles. 

In an intermediate regime, the gap betweeen the coreshell modes $\Omega_\textrm{cs}$ will still be much larger than the gaps open by the alternation of distances in the SSH chain, $\Omega_{\pm,\mu}$. This means that as long as $\gamma \ll \Omega_\textrm{cs}$, the bonding mode of one coreshell particle in the dimer with the antibonding mode of the other particle will be deeply off-resonance so these couplings will be negligible. We will have 3 pairs of bands centered around each of the single coreshell particle resonance frequencies, one per polarization.

When $|1-\beta|\rightarrow 1$, the bulk modes tend to the frequencies of isolated dimers, this is $G_2 = 0 \; (\beta = 0)$ or $G_1 = 0 \; (\beta = 2)$. However, these extreme cases are outside of the accuracy limit of the dipole approximation which is accurate only for inter-particle gaps of at least the radius of the particles. 

In Figure~\ref{fig:dispersion_SSH} we plot the dispersion bands around $\hbar\omega_{cs}^+  = 3.71$eV (panel (a)) and   $\hbar\omega_{cs}^-  = 1.95$eV (panel (b)), for the same set of parameters as the previous figures, with period and spacing of the chain $d = 200$nm and  $\beta = 1.2$. The purple and gray lines correspond to transversal and longitudinal modes, respectively. The $yy$ and $xx$ modes are degenerated due to the isotropy in the transversal plane, so orange bands are doubly degenerate. We represent the dipoles of each coreshell particle as black arrows. Each dipole can be decomposed as the sum of the dipoles of the sphere and the void modes, so each coreshell SSH mode can be seen as the hybridization between four dipoles. 

The gaps between the bands can be derived from Equation~\eqref{eq:dispersion_bands}, as
\begin{eqnarray}
\Omega_{\pm,\mu} = |\omega_{\mu,\pm,2}(\pm \pi/d) - \omega_{\mu,\pm,1}(\pm \pi/d)|
\end{eqnarray}
For this set of parameters we find $\hbar\Omega_{x,+} = 0.044~$eV, $\hbar\Omega_{y,+} =  \hbar\Omega_{z,+}= 0.022~$eV, $\hbar\Omega_{x,-} = 0.028~$eV, $\hbar\Omega_{y,-} =\hbar\Omega_{z,-}= 0.014~$eV. We see these splittings are small compared to the gap between the modes of the single coreshell nanoparticle, $\hbar\Delta\omega_\textrm{cs} = 1.76~$eV, as expected. The gaps between the pairs of bands must be proportional to $|g(\pm \pi)| = |G_1 - G_2|$. When $\beta = 1$, $G_1 = G_2$ and the bands are ungapped, so the only gap is the one coming from the hybridization in the single coreshell nanoparticle. 
\begin{figure}[!h]
    \centering
    \includegraphics[width=0.99\linewidth]{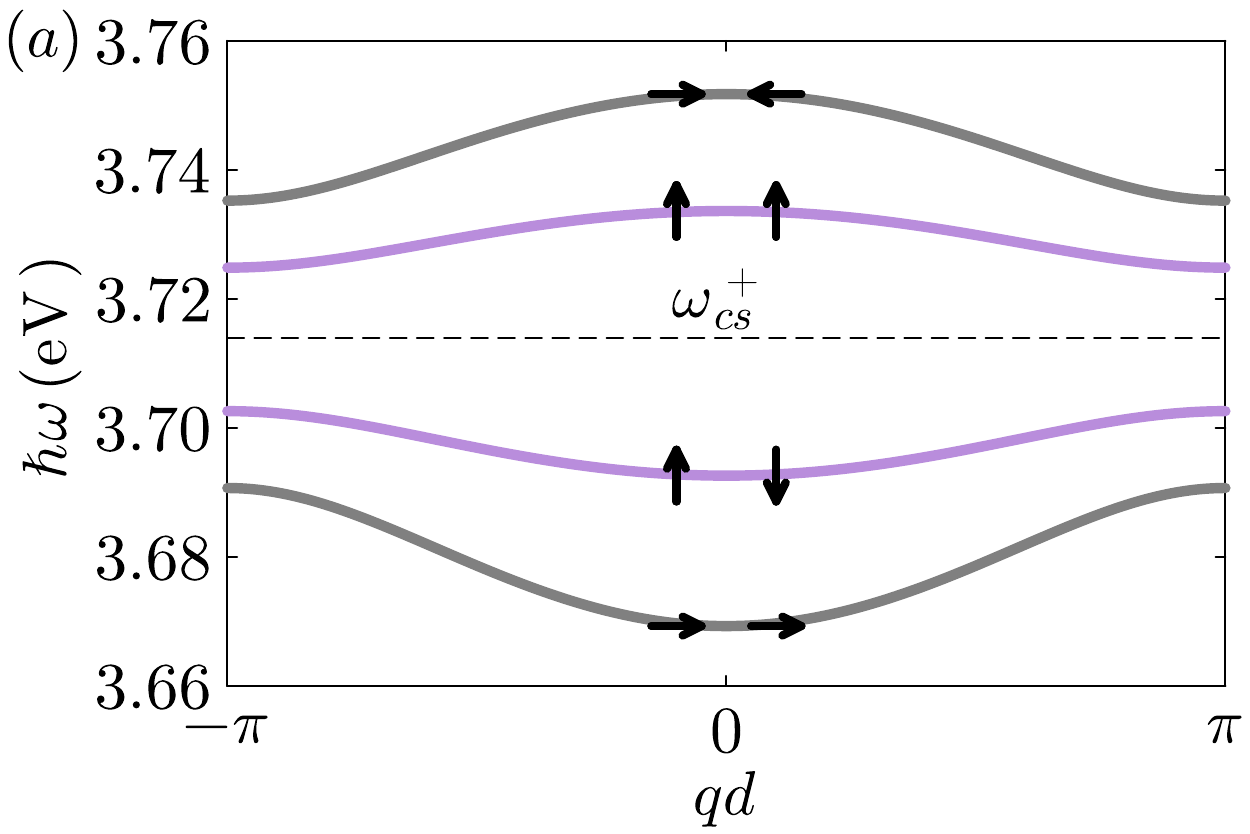}
        \includegraphics[width=0.99\linewidth]{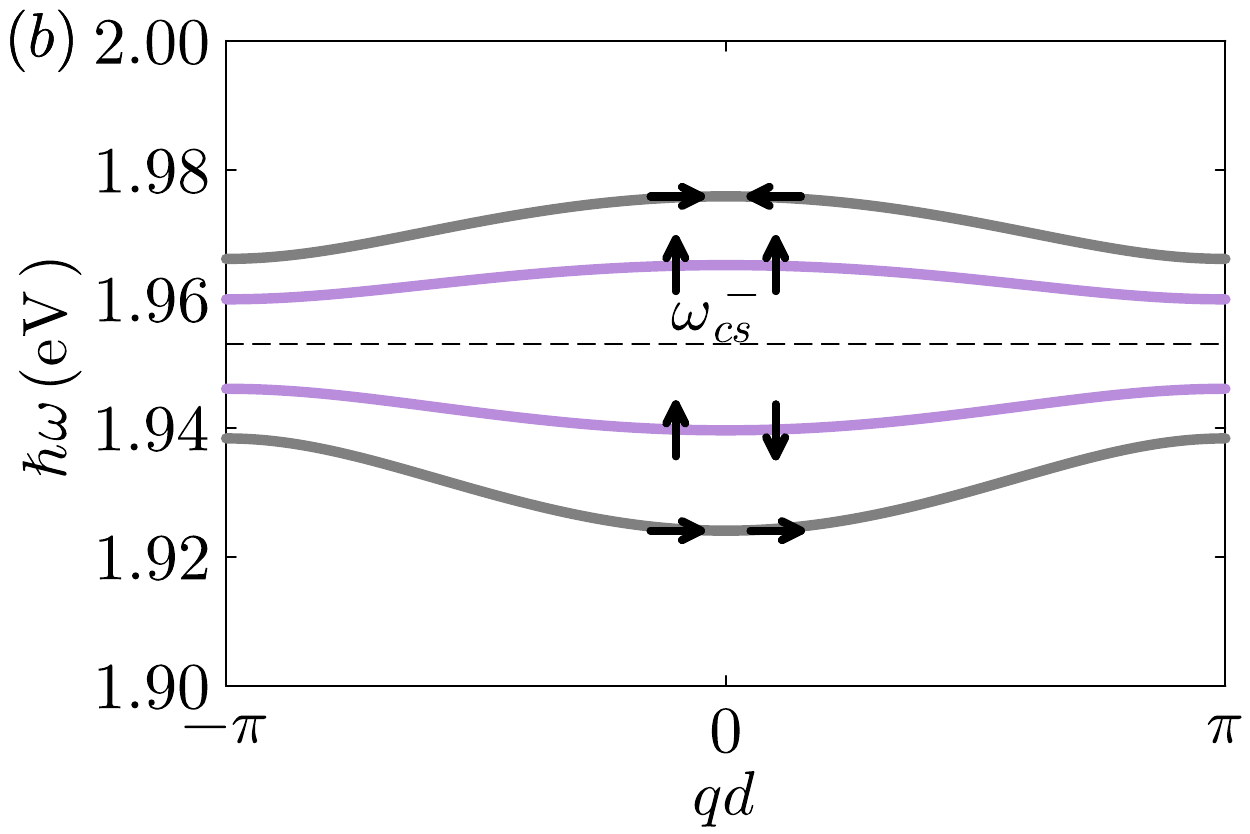}
    \caption{Dispersion bands of SSH: In panel (a) we represent the bands around the upper coreshell resonance frequency $\hbar\omega_\textrm{cs}^+ \simeq 3.71$eV and in panel (b) the same for the lower resonance frequency $\omega_\textrm{cs}^- \simeq 1.95$eV. Purple and gray lines represent the dispersion bands for transversal and longitudinal modes. The parameters used are the same as in previous figures and $d = 200 $nm and $\beta = 1.2$. }
    \label{fig:dispersion_SSH}
\end{figure}
\subsection{Topological invariants}

Topological systems can be classified by their symmetries and characterized by topological invariants \cite{Chiu2016}. In the case of the SSH model, the topological invariants are related to the number or existence of edge states. Therefore, the edge states, which only manifest in the finite array, can be predicted from the periodic array. This is known as bulk-boundary correspondence \cite{Asboth2016}.

The SSH model can be caracterized either by the winding number $W$ or by the Zak phase, $Z$. We will show both invariants are equivalent and related by $Z = \pi W \mod{2\pi}$. 

As we commented previously, the coreshell array can be mapped to a bilayer SSH. The topological invariants and phases of multiband SSH models have been studied in Reference~\cite{Lee2022}. Here, first we calculate the Zak phase, which is a 1D version of the Berry's phase, and is the phase picked by the eigenmode of the band under the Fermi energy crossing the Brillouin zone. In the plasmonic SSH, the Zak phase can be derived from the dipolar momentum eigenvectors $\mathcal{P}(q)$ \cite{Pocock2018,Pocock2019}. However, in a plasmonic system there is no Fermi energy, so all the bands are equally relevant. Due to this, it is more useful to define a Zak phase of each band,
\begin{align}
    Z_{\mu,\pm,n} &=i\int_{-\pi/d}^{\pi/d} \mathcal{P}^\dag_{\mu,\pm,n}(q) \frac{\partial \mathcal{P}_{\mu,\pm,n}(q)}{\partial_q}   dq 
\label{eq:Zak_plasmonic} 
\end{align}
As we see in  Equation \eqref{eq:G_matrix_periodic} all the dependence in $q$ is in the sublattice matrix, $\mathcal{G}_0(q)$. Consequently, as long as we do not break the internal spherical symmetry of the coreshell nanoparticle and we don't have cross-polarization couplings, all the topological non-triviality of the Green's matrix will be enclosed in this matrix. 

Then, the Zak phase can be reduced to
\begin{align}
Z_{\mu,\pm,n} &=i\int_{-\pi/d}^{\pi/d} (\mathcal{P}_{0}^n(q))^\dag\frac{\partial \mathcal{P}_{0}^{n}(q)}{\partial_q}   dq \nonumber \\ &= \frac{\phi_{n}(q=\pi/d) - \phi_{n}(q=-\pi/d)}{2} \operatorname{mod} 2\pi,
\end{align}
where $\mathcal{P}_0^{n}$ is the $n$th eigenmode of $\mathcal{G}_0(q)$, which can be defined in terms of a relative phase, $\phi_n(q)$,
\begin{eqnarray}
\mathcal{P}_0^n(q) = \begin{pmatrix} e^{-i\phi_n(q)} \\ 1 \end{pmatrix},
\end{eqnarray}
We find all the bands have the same Zak phase. In Figure~\ref{fig:topological_invariants}(a) we plot the Zak phase $Z$ normalized by $\pi$ depending on $\beta$. We see the Zak phase is 0 for $\beta <1$, and 1 for $\beta >1$, signaling a topological transition at $\beta = 1$. 

Another topological invariant usually used to characterize the SSH model is the winding number, W. $\mathcal{G}_0(q)$, as any $2\times 2$ matrix,  can be defined in terms of the Pauli matrices $\sigma_{1,2,3}$ and the $2\times 2$ identity matrix, $\mathbb{I}_2$. 
\begin{eqnarray}
\mathcal{G}_{0}(q) = \textbf{g}(q) \cdot \boldsymbol{\sigma}= &g_0(q) \mathbb{I}_2+ g_1(q) \sigma_1 \nonumber \\ + &g_2(q) \sigma_2 +  g_3(q) \sigma_3.
\end{eqnarray}
In this case, $g_0(q) = g_3(q) = 0$ and $g_1(q) = G_1 - G_2\cos{q}$ and $g_2(q) = -G_2\sin{q}$.  
\begin{figure}[!h]
    \centering
       \includegraphics[width=0.99\linewidth]{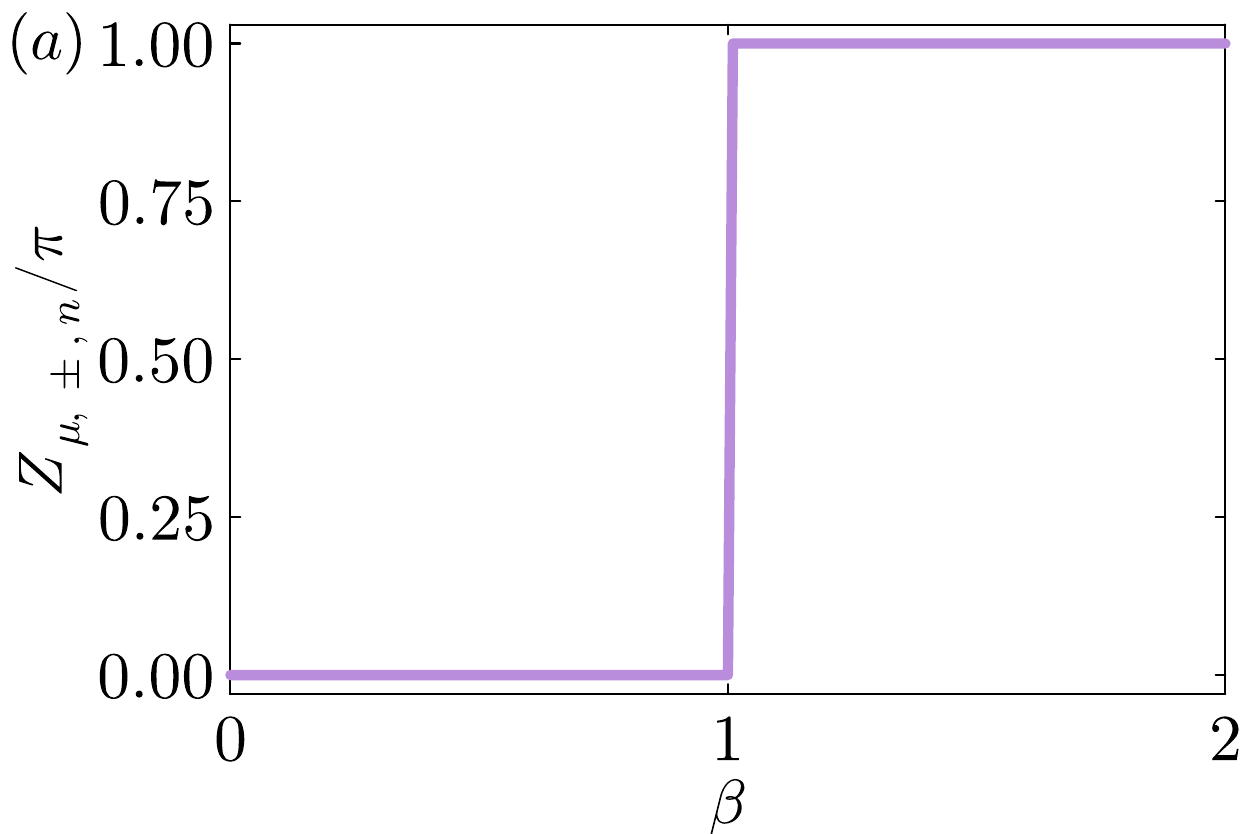}
    \includegraphics[width=0.99\linewidth]{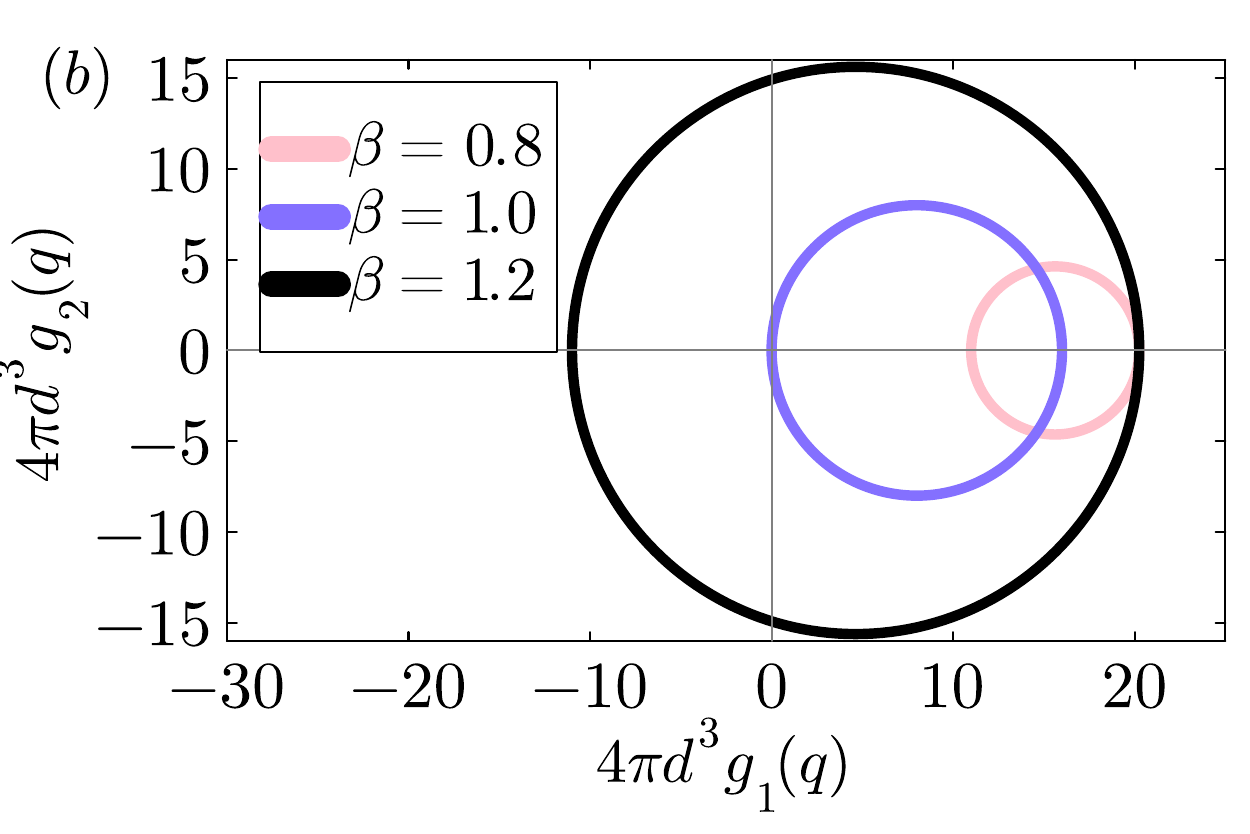}
    \caption{Topological invariants. (a) Zak phase of the dispersion bands of the SSH chain, depending on $\beta$. The jump from 0 to 1 at $\beta = 1$ signals the topological transition. We used the same parameters as in the previous figures. (b) Winding number $W$ of $\mathcal{G}_0(q)$ for $\beta = 0.8,  1, 1.2$. The rest of parameters is the same as in previous figures. We plot the trajectories $[g_1(q), g_2(q)]$. For $\beta < 1$ (pink line), the trajectory don't enclose the origin, so the winding number is $W=0$, indicating a trivial phase. At $\beta = 1$ (blue line), the trajectory touches the origin. For $\beta > 1$ (black line), the trajectory winds once around the origin, $W=1$, indicating a non-trivial phase.}
    \label{fig:topological_invariants}
\end{figure}

The winding number is defined as the number of times the function $[g_1(q), g_2(q)]$ winds around the origin.  In Figure~\ref{fig:topological_invariants}(b), we plot the Pauli trajectories depending on $\beta$. First, we plot the $\beta = 0.8$ (pink line), and we see this trajectory doesn't enclose the origin, so the winding number is $W = 0$. For $\beta = 1$ (blue line), the trajectory intersects the origin. Finally, for $\beta = 1.2$ (black line), the trajectory winds once around the origin, so $W=1$. Therefore, a topological transition occurs at $\beta = 1$, going from $W = 0$ for $\beta < 1$ to $W=1$ for $\beta > 1$. 
\subsection{Sublattice symmetries}

Sublattice symmetry represents the existence of two identical sublattices which are only coupled between each other, as the first-neighbours SSH model. 

It can formally be defined as \cite{Asboth2016}: 
\begin{eqnarray}
\mathcal{G}(q) = -S^{\dag}\mathcal{G}(q) S \label{eq:SLS}.
\end{eqnarray}
where $S$ is the sublattice inversion operator,
\begin{eqnarray}
S = \sigma_3 \otimes \mathbb{I}_2 \otimes \mathbb{I}_3   =  \begin{pmatrix} 1 & 0 \\ 0 & -1 \end{pmatrix} \otimes \mathbb{I}_6.
\end{eqnarray}

In terms of Pauli matrix decomposition of $\mathcal{G}_0(q)$, this implies that $g_0(q) = g_3(q) = 0$. As long as the particles are identical the polarizability $A_{0}(\omega)$ term breaks the sublattice symmetry in a trivial manner, shifting the eigenvalues while leaving invariant the eigenmodes, so we refer to this as trivial symmetry breaking \cite{Pocock2018,Pocock2019}.

The sublattice symmetry is the symmetry which protects the topological edge states. Once this symmetry is broken, the edge states are not fixed to $\omega_{sp}$ and for strong perturbations, they could be pushed out of the gap and hybridize with the bulk states.

By looking at Equation~\eqref{eq:Gmatrix_fin}, we can see the polarization and the different surface modes also act like extra subspaces or sublattices. These degrees of freedom could be exploited to induce non-trivial topology similarly to the sublattices of the SSH model. 

The hybridization between the longitudinal and transversal modes has also proven effective to induce non-trivial phases, either by breaking the 1D nature of the lattice as in zigzag chains \cite{Poddubny2014}, or by orienting non-spherical nanoparticles \cite{Buendia2023}. 

Analogously, breaking the internal spherical symmetry of the coreshell could lead to non-triviality. When the cells are non-concentric, the dipoles of the inner and outer modes would not be at the same position, which would translate into an alternation between the couplings similar to the one from the SSH model. In Appendix~\ref{sec:cs_topology} we briefly consider this symmetry breaking. 
\subsection{Spectrum and edge states}
\label{subsec:spec}
\begin{figure}[!h]
    \centering
    \includegraphics[width=0.91\linewidth]{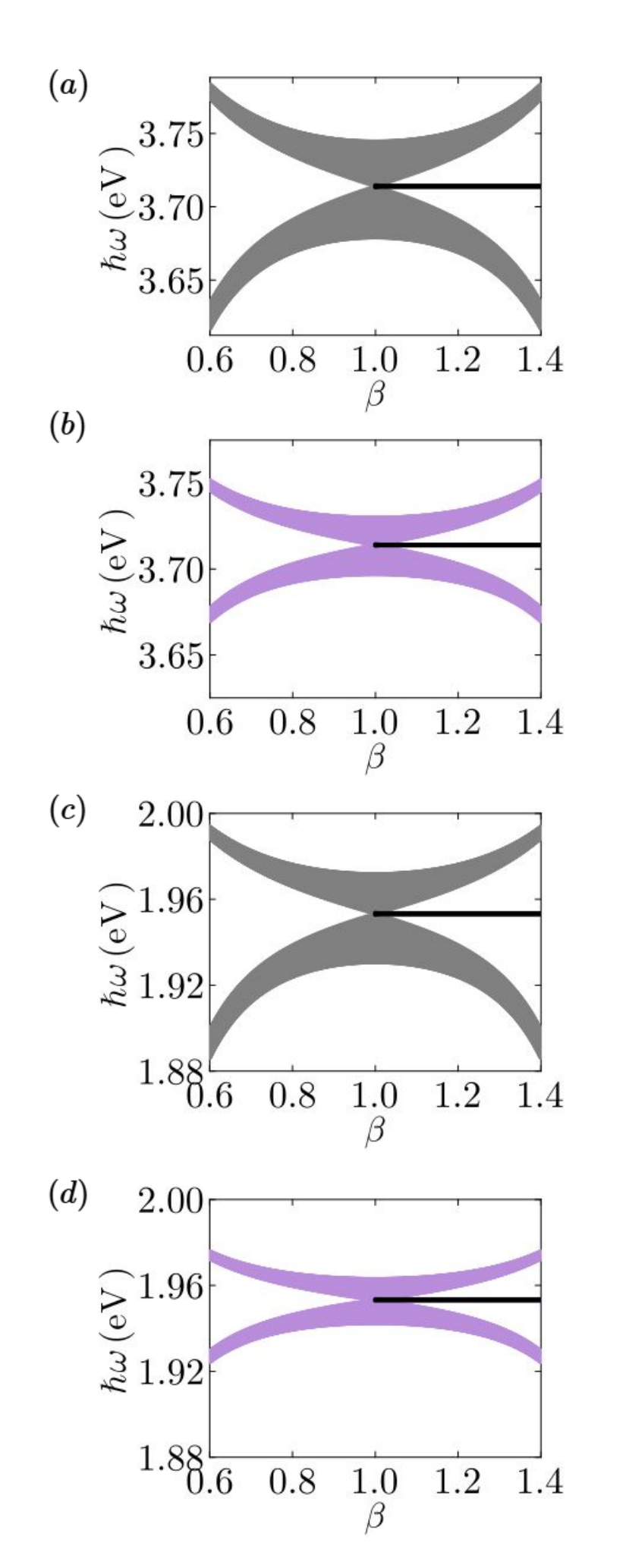}
    \label{fig:spectrum_SSH}
    \caption{Spectrum of a SSH chain of coreshell nanoparticles, depending on $\beta$. Gray and purple lines represent bulk modes while black lines represent edge states. (a) Longitudinal modes from upper gap. The edge states arise after the topological transition, for $\beta > 1$. (b) Transversal mode for upper gap.  (c) Longitudinal modes from lower gap (d) Transversal modes from lower gap. Same parameters as previous figures.}
\end{figure}
\begin{figure}[!h]
    \centering
    \includegraphics[width=0.99\linewidth]{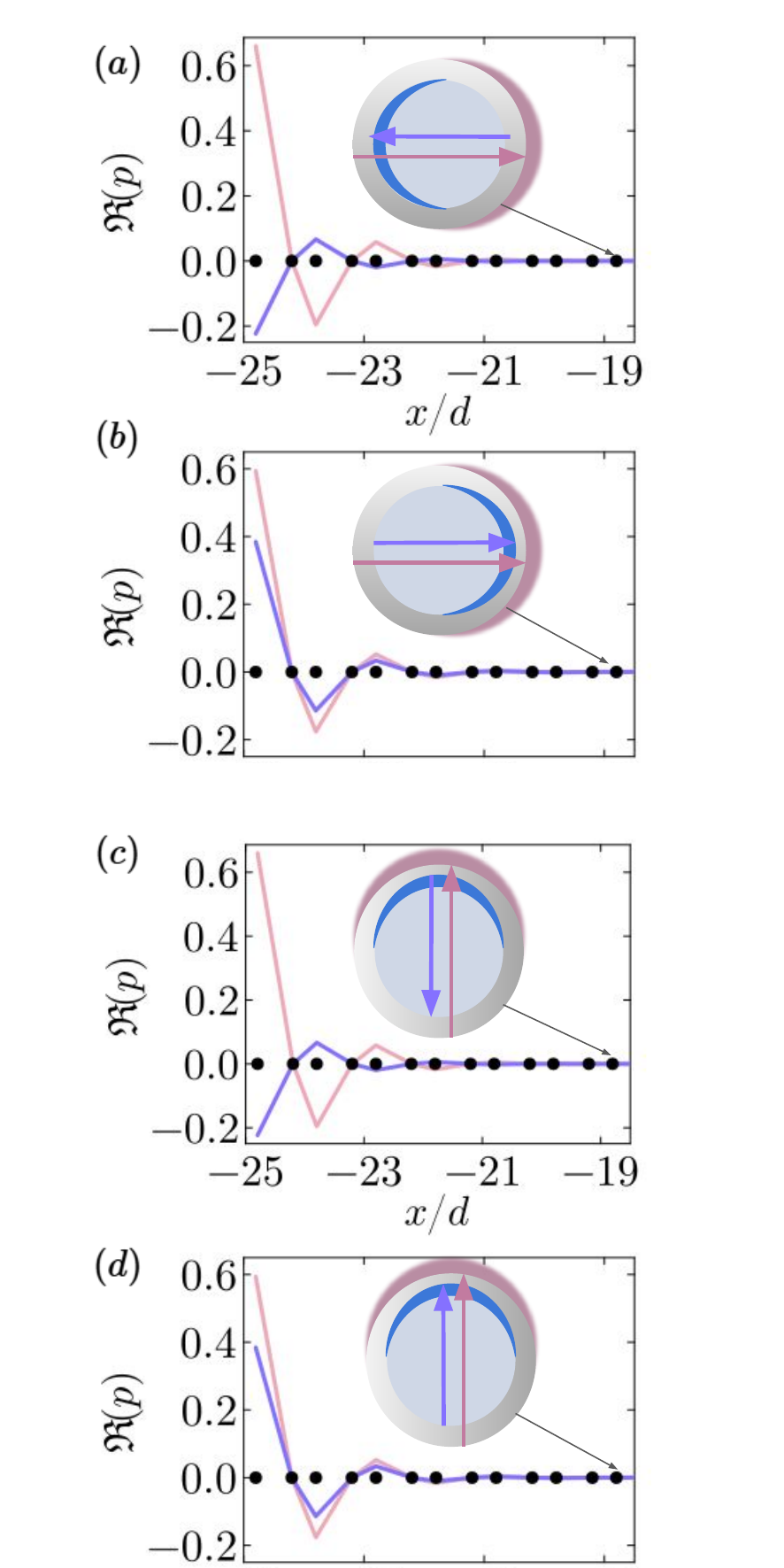}

    \caption{Edge states of a coreshell SSH chain of coreshell nanoparticles: Zoom of the edge state eigenmodes of the left ends of the chain for the upper and lower gaps for (a) Longitudinal polarization and upper gap (b)  Longitudinal polarization and lower gap (c) Transversal polarization and upper gap (d) Transversal polarization and lower gap. The pink and blue lines represent the dipolar momentums of the inner and outer surface modes $p_a$ and $p_b$ for the edge state eigenmode. The black dots represent the sites in the lattice. We used the same set of parameters as in previous figures, for $\beta = 1.2$ and $n_c = 50$ unit cells.}
    \label{fig:edge_states_SSH}
\end{figure}
The spectrum of the finite array can be calculated as
\begin{eqnarray}
\lambda_n(\mathbb{A}^{-1}(\omega) - \mathbb{G}) = 0.
\end{eqnarray}
Differently from the infinite array, the analysis of the resonant frequencies of the finite lattice involves large matrices, which makes impossible to solve it analytically. Therefore, we have to solve it numerically. The coupled-dipole decomposition for the nanoparticles is more useful in the finite array, as it allows to calculate each mode independently, without having to finetune the starting frequency for the solver to find the modes.

Solving for each eigenvalue, we can find the resonant frequencies of the $3n_\textrm{res} n_c\mathcal{N}$ modes. We plot the spectrum for $n_c=50$ unit cells depending on $\beta$.  In panel (a,b) we plot the spectrum around the upper resonance frequency $\hbar\omega_\textrm{cs}^{+} = 3.71~$eV and in panel (c,d) we plot the spectrum around $\hbar\omega_\textrm{cs}^- = 1.95~$eV. We plot the bulk longitudinal modes in gray panel (a,c) and the bulk transversal modes (b,d) in purple, with the edge states as black lines.

We see the spectrums around both of the resonances are almost identical, except for the gap, which is larger for the higher frequency modes. At $\beta = 1$ the gaps close. As in the periodic system, the spectrum is symmetrical with respect to $\beta = 1$ for the bulk modes. However, after the transition, for $\beta >1$, the array hosts topological edge states, localized on their ends. The topological invariants represent the number of pair of edge states in the system, due to the bulk-boundary correspondence. 

We see that even when for $\beta \gtrsim 1$ there is a small splitting between longitudinal and transversal modes, the edge states tend to the resonant frequencies of the single coreshell particle, $\omega_\textrm{cs}^{\pm}$. Due to the mirror symmetry between the left and right ends of the chain and the isotropy of the particles, the edge states have a sextuple degeneracy. The degeneracy between the edge states of different polarizations can be broken either by employing anisotropic materials or by using non-spherical, e.g. spheroidal, particles \cite{Proctor2020,Buendia2023}. 

As the frequencies of the upper and lower frequency edge states can be tuned by the background permittivity and the filling ratio of the particles, this could be employed for topologically-protected generation of higher harmonics. This have been proposed with Mie modes in edge states of zigzag chains of dielectric particles \cite{Kruk2019}. High armonic generation by single coreshell nanoparticles has also been studied in Reference \cite{Scherbak2018}. 

In Figure \ref{fig:edge_states_SSH}, we plot the eigenmodes of $(\mathbb{A}^{-1}(\omega) -\mathbb{G})$ corresponding to (a) longitudinal mode and upper gap (b) transversal mode and upper gap (c) longitudinal mode and lower gap and (d) transversal mode and lower gap.  The difference between the edge states around $\omega_\textrm{cs}^+$ and $\omega_\textrm{cs}^-$ is in the relative sign between the dipoles within the coreshell nanoparticle, positive for the bonding mode and negative for the antibonding mode. 

We find the eigenmodes are also independent of polarization, which suggests the polarizability term only affects the amplitude of the gaps and the relative factors between the dipoles of the cavity and sphere modes, but not the decay constant of the edge states.

\section{Conclusions}
Coreshell nanoparticles are an interesting platform for optical applications, combining functionalities from different materials. The optical response of coreshell nanoparticles can be understood via hybridization models, which analyzes the resonant modes of structured metallic particles in terms of the coupling of modes of simpler nanostructures. Specifically, the resonant modes of a coreshell nanoparticle can be seen as the hybridization of the surface modes of a metallic sphere and a metallic void. This coupling leads to two modes, the bonding and antibonding, depending on if the dipoles are parallel or antiparallel. 

Here we propose to combine the hybridization theory with the coupled electric dipole formalism to study periodic arrays of coreshell nanoparticles. By making this, we propose a formalism where the number of bands matches the number of resonant modes, allowing to study topological properties of the bands. We see that the resulting system can be mapped to multilayer SSH arrays. In this frame, the different surface modes now act like an extra sublattice or subspace, which can be exploited to tune the optical and topological properties of the system. 

Within this model, we analyze analytically and numerically a Su-Schrieefer-Hegger chain of Si@Ag coreshell nanoparticles. We show the two resonances coming from the layered particle leads to the existence of two topological gaps at different frequencies for a SSH chain, with one pair of edge states per polarization and per gap. We show this can be mapped to a multilayer SSH chain. The multigap edge states could be employed for non-linear applications, as their frequencies can be tuned by the filling factor of the coreshell nanoparticle and the host permittivity.

\section*{Acknowledgement}
A.B. and N. M. R. P. acknowledges support from
the European Union through the EIC PATHFINDER
OPEN project No. 101129661-ADAPTATION. N.M.R.P. acknowledges support from
the Portuguese Foundation for Science and Technology (FCT) in the framework of the Strategic Funding UIDB/04650/2020, COMPETE 2020, PORTUGAL 2020, FEDER, and through project PTDC/FISMAC/2045/2021. A.B. thanks Marcelo Barreiro for fruitful discussions.

\appendix 
\counterwithin{figure}{section}
\section{Coreshell dipolar decomposition}
\label{sec:coreshell_decomp}
As we explained in the main text, the optical resonance modes of layered nanoparticles can be understood as the hybridization of the surface modes at the different interfaces. 

Therefore, for a coreshell nanoparticle with $n_l$ layers and $n_\textrm{int} = n_l-1$ interfaces, supposing the medium of each layer is either non-resonant or have only a resonance, the number of resonances is:
\begin{eqnarray}
n_\textrm{res} \leq 2 n_\textrm{int} = 2(n_{l}-1).
\end{eqnarray}
TThe modes of a nanoshell can be seen as the hybridization of the void and sphere surface modes at the multiple interfaces. The resonance frequencies are the solutions of the Fröhlich condition, $\Re(\varepsilon_1(\omega))+2\Re(\varepsilon_2(\omega)) = 0$. 
Therefore, we can decompose an $n_{l}$ particle into $2(n_{l}-1)$ simple nanostructures with only one resonance coupled to each other. In other words, we can take a single coreshell nanoparticle with concentric shells as a system of $n_\textrm{res}$ coupled dipoles placed at the same position. This is different from the discrete dipole approximation (DDA), where a single particle is decomposed as a 3D array of polarizable cubes \cite{Yurkin2007}. In our model, a polarizability with $n_\textrm{res}$ poles would be represented by a $3n_\textrm{res}\times 3n_\textrm{res}$ matrix:
\begin{eqnarray}
&\overleftrightarrow{A}_\textrm{cs}(\omega) = \nonumber \\&\begin{pmatrix} \overleftrightarrow{\alpha}_1^{-1}(\omega) & \overleftrightarrow{g}_{12} & \cdots & \overleftrightarrow{g}_{1n_\textrm{res}} \\  \overleftrightarrow{g}_{21} & \overleftrightarrow{\alpha}_2^{-1} & \cdots & \overleftrightarrow{g}_{2n_\textrm{res}}(\omega) \\ \vdots & \vdots & \ddots & \vdots \\ \overleftrightarrow{g}_{n_\textrm{res}1} & \overleftrightarrow{g}_{n_\textrm{res}2}  & \cdots & \overleftrightarrow{\alpha}_n^{-1}(\omega) \end{pmatrix}^{-1}
\label{eq:coupled_pol_cs}
\end{eqnarray}
The polarizability of the coreshell in the single-dipole and the coupled-dipole descriptions are related by
\begin{eqnarray}
\alpha_\textrm{cs}(\omega) = \begin{pmatrix} 1 & \cdots & 1 \end{pmatrix} A_{\textrm{cs}}(\omega) \begin{pmatrix} 1 \\ \vdots \\ 1 \end{pmatrix}.
\end{eqnarray}
The analytical single-dipole polarizability of a coreshell nanoparticle with several shells can be obtained by the iterative method proposed by Reference \cite{Ugwuoke2024}. The derivation of the effective couplings in Equation~\eqref{eq:coupled_pol_cs} is not trivial, and need a case by case study.  

For the coreshell SSH array we saw that we can map the to multilayer tight binding SSH arrays. So generally, by making this decomposition any monolayer array of coreshell nanoparticles can be mapped to a tight binding model of a multilayer lattice with $n_\textrm{res}$ layers.

\section{Pseudo-sublattice-symmetry and coreshell-induced topological phase}
\label{sec:cs_topology}

In order to study possible non-trivial phases coming from the hybridization between the states of different interfaces in coreshell nanoparticles, we imagine now a 1D Bravais chain. This is, a chain with periodicity $d$ and one particle per unit cell instead of two like the SSH chain from the main text. 
\begin{figure}[!h]
    \centering
\includegraphics[width=1.0\linewidth]{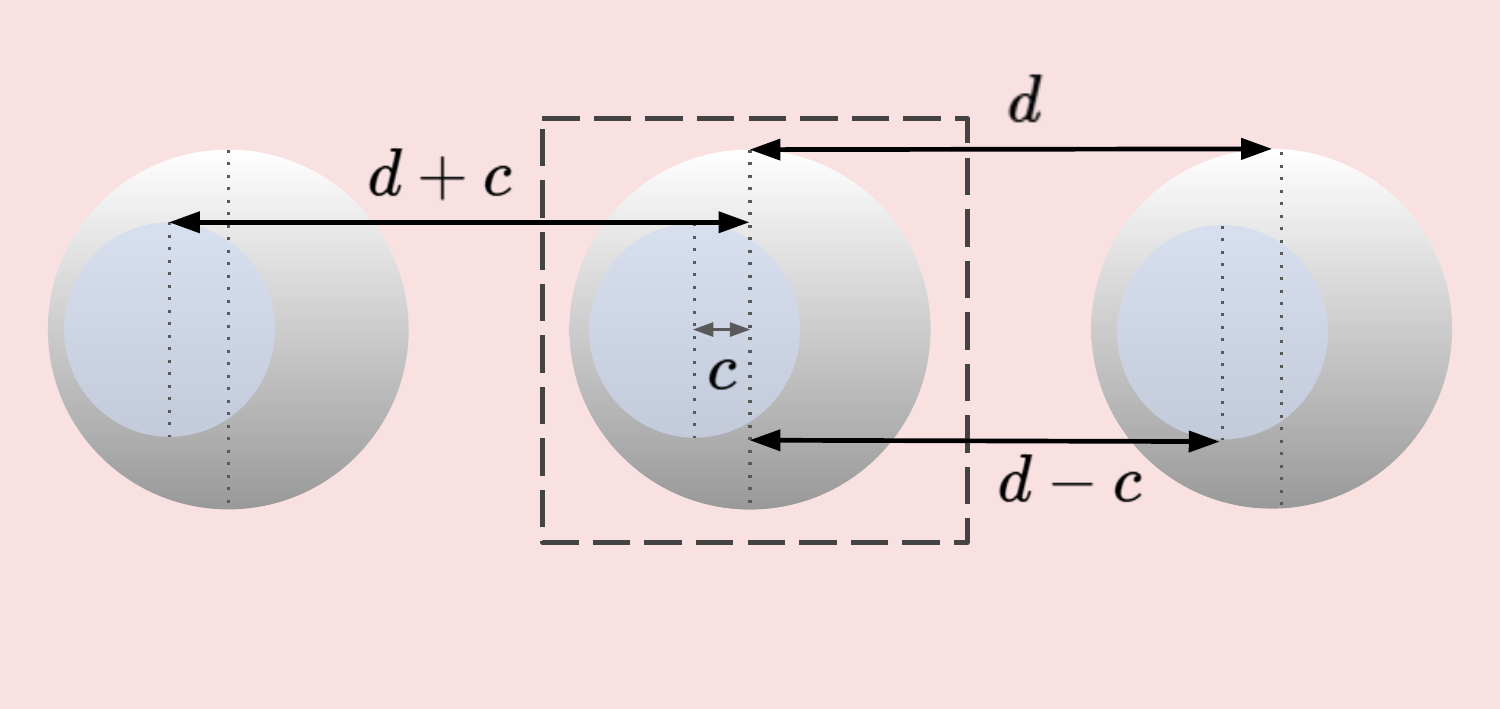}
    \label{fig:non-concentric-cs}
    \caption{1D Bravais chain of non-concentric coreshell nanoparticles. The core is displaced  a distance $c$ in the array axis. By breaking the internal symmetry of the coreshell nanoparticles we get different couplings between the core and the shell from adjoint cells, similarly to the alternating coupling strengths in the plasmonic SSH chain.}
\end{figure}
We can define an alternative sublattice symmetry, 
\begin{eqnarray}
(\mathcal{A}^{-1}(\omega) -\mathcal{G}(q))  = -\widetilde{S}^\dag(\mathcal{A}^{-1}(\omega) - \mathcal{G}(q))\widetilde{S} 
\end{eqnarray}
where now the sublattice operator, $\widetilde{S} = \mathbb{I}_3\otimes \sigma_3$, acts on the "shell" space. 

For the concentric coreshell nanoparticle, we have now, 
\begin{eqnarray}
&\mathcal{A}^{-1}(\omega) - \mathcal{G}(q) = \mathbb{A}_\textrm{cs}^{-1}(\omega) \otimes \mathbb{I}_3   \nonumber \\ &-\begin{pmatrix} 2G_d\cos(q) & 2G_d\cos(q) \\ 2G_d\cos(q) & 2G_d\cos(q) \end{pmatrix} \otimes \mathbb{M}_\textrm{pol}(\hat{\textbf{x}})
\end{eqnarray}
So for each polarization, we have: 
\small
\begin{eqnarray}
\mathbb{A}_{cs}^{-1}(\omega) - m_\mu\mathcal{G}_{s}(q) \nonumber = \\ \begin{pmatrix} \widetilde{\alpha}^{-1}_a(\omega) - 2m_\mu G_d\cos(q) & -2m_\mu G_d\cos(q) + g_{ab} \\ -2m_\mu G_d\cos(q) - g_{ab} & \widetilde{\alpha}_b^{-1}(\omega)-2m_\mu G_d\cos(q) \end{pmatrix}.
\end{eqnarray}
\normalsize
The $g_1(q) =  -2m_\mu G_d\cos(q)$ and $g_2(q) = -i g_{ab}$. As we see, as the second term does not depend on $q$, the trajectory is flat, which means the winding number is always zero. 

In the SSH model, the topological bandgap opening is produced by the breaking of the mirror symmetry with respect to the atom. Analogously, in order to achieve non-trivial phases arising from the shell hybridization we need to break some symmetries within the particle. We now imagine the core is displaced in the direction of the axis of the chain a distance $c$ from the nanoparticle center. We can see a similarity of this system with the nanoparticle dimers in the SSH chain. Now the hybridization between the surfaces can be as two coupled dipoles at different positions. We do not calculate the explicit expression for the effective coupling between the surface modes in this geometry and just leave it as $g_{ab}$. The polarizability of non-concentric coreshell nanoparticle can be found in Reference \cite{Zhang2013}.

Now the matrix $\mathbb{A}_\textrm{cs}^{-1}(\omega) - m_\mu\mathcal{G}_{s}(q)$ is
\small
\begin{eqnarray}
\mathbb{A}_\textrm{cs}^{-1}(\omega) - m_\mu\mathcal{G}_{s}(q) = \nonumber \\ \begin{pmatrix}\widetilde{\alpha}^{-1}_a(\omega) - 2m_\mu G_d\cos(q)& G_{d-c}e^{-iq} + G_{d+c}e^{iq} + g_{ab} \\  G_{d-c}e^{iq} + G_{d+c}e^{-iq} - g_{ab} & \widetilde{\alpha}^{-1}_b(\omega) - 2m_\mu +G_d\cos(q) \, \end{pmatrix}.
\end{eqnarray}
\normalsize

Ignoring the diagonal terms of the matrix, we focus on $g_1(q) = (G_{d-c} + G_{d-c}) \cos(q) + i \Re(g_{ab}) $ and $g_2(q) = (G_{d+c} - G_{d-c})  \sin(q) + \Im(g_{ab})$. Ignoring the imaginary component of $g_1(q)$, we see this is an ellipse of semi-axes $G_{d+c} + G_{d-c}$ and $G_{d-c} + G_{d-c}$ centered at $(0,\Im(g_{ab}))$. For a non-Hermitian sublattice-symmetric matrix, the winding number \cite{Yin2018}, can be expressed on terms of two winding numbers around the points $P_1 = (\Im(g_2), -\Im(g_1))$ and $P_2 = (-\Im(g_2), \Im(g_1))$, which in this case are $P_1 = (0,\Re(g_{ab}))$ and $P_2 = (0, -\Re(g_{ab}))$. The total winding number is $W = \frac{W_1 + W_2}{2}$. The Pauli trajectory in this case winds around both points when  $(G_{d-c} + G_{d-c}) > \Re(g_{ab})$ and $(G_{d+c} - G_{d-c}) > \Im(g_{ab})$. So we would have a non-trivial phase in this regime. 

However, the diagonal terms break the sublattice symmetry. We have a diagonal component proportional to the identity, $g_0(q) = 2G_dcos(q) + \frac{(\widetilde{\alpha}^{-1}_a(\omega) + \widetilde{\alpha}^{-1}_b(\omega))}{2}$ which produces a trivial breaking of the sublattice-symmetry. On the other hand, the $g_3(q) = \frac{(\widetilde{\alpha}^{-1}_a(\omega) - \widetilde{\alpha}^{-1}_b(\omega))}{2}$ term breaks the sublattice symmetry in a non-trivial matter, which compromises the topological protection of the potential edge states. This non-trivial sublattice-symmetry breaking can be minimized by maximizing the spectral overlap between void and sphere modes.
\section{Retarded and radiating SSH chain of coreshell nanoparticles}
\label{sec:long-range}
When particle size is not negligible compared to the wavelength of light, the quasi-static approximation looses its acuracy. For larger particles, $a\gtrsim 20$nm, we need to make corrections to the quasi-static (also known as long-wavelength) polarizability \cite{Moroz2009}. While the exact polarizability for larger particles is given by Mie theory, there is an ad-hoc intermediate correction known as modified long wavelength approximation (MLWA). The MLWA corrections for the polarizability of the sphere and cavity modes are \cite{Moroz2009}
\begin{eqnarray}
\alpha_a^\textrm{MLWA}(\omega) =  \frac{\alpha_a(\omega)}{1-\frac{k^2}{a}\alpha_a(\omega) -\frac{2i k^3}{3}\alpha_a(\omega)} \nonumber \\ \alpha_b^\textrm{MLWA}(\omega) =  \frac{\alpha_b(\omega)}{1-\frac{k^2}{b}\alpha_b(\omega) -\frac{2i k^3}{3}\alpha_b(\omega)} 
\end{eqnarray}
where the second term in the denominator accounts for the optical damping, producing a spectral broadening, and the last one accounts for retardation and produces a size-dependent red-shift of the LSPR. 
We see that when $kb \ll 1$ and $ka\ll 1 $, these corrections are negligible and the MLWA polarizabilities tend to the quasistatic polarizabilities. 

The MLWA correction for the polarizability of the coreshell nanoparticle is~\cite{Chung2009},
\begin{eqnarray}
&\alpha_{\mathrm{cs}}^{MLWA}(\omega) = \frac{4}{3}\pi\varepsilon_0 a^3 \cdot  \nonumber \\ &\frac{(\varepsilon_c q_b +\varepsilon(q_b-1))(\varepsilon-\varepsilon_B) - \frac{b^3}{a^3}(\varepsilon_c-\varepsilon)(\varepsilon(q_a-1)+\varepsilon_B q_a)}{(\varepsilon q_a - \varepsilon_B(q_a-1))(\varepsilon_c q_b - \varepsilon(q_b-1)) - \frac{b^3}{a^3}(\varepsilon-\varepsilon_B)(\varepsilon_c-\varepsilon)q_a(q_a-1)},
\label{eq:alphamlwa_cs}
\end{eqnarray}
where $q_a = \frac{1}{3} - \frac{k^2 a^2}{3} - i\frac{2k^3 a^3}{9}$ and $q_b = \frac{1}{3} - \frac{k^2 b^2}{3} - i\frac{2k^3 b^3}{9}$. When $ka \ll 1 $ and $kb \ll 1$, we recover the quasistatic polarizability of the coreshell nanoparticle. As the corrections are proportional to $k$ and therefore to the frequency, the lower frequency mode of the coreshell nanoparticle, $\omega_{cs}^-$ will be less affected than the upper mode, as we see in Reference \cite{Chung2009}.

Now we consider the correction to the Green's dipole-dipole function. In this case, when we are not in the regime $kd \ll 1$, we need to include the medium and long-range terms. Additionally, the first-neighbours approximation is not enough anymore, so we need to consider the interaction between all the particles. When the two dipoles are separated  in the $x$-axis, the Green's dipole-dipole function is:
\begin{eqnarray}
&\GG(\omega, R\hat{\mathbf{x}}) = \frac{e^{ikR}}{4\pi R}\bigg[\left(1 + \frac{i}{kR} - \frac{1}{k^2 R^2}\right)\mathbb{I}_3 \nonumber \\ &- \left(1 + \frac{3i}{kR} -\frac{3}{k^2 R^2}\right)\hat{\mathbf{x}}\otimes \hat{\mathbf{x}} \bigg] = \nonumber \\ &=\begin{pmatrix} 2 & 0 & 0 \\ 0 & -1 & 0 \\ 0 & 0 & -1 \end{pmatrix}\left(G_{SR}(\omega,R) +G_{MR}(\omega,R)\right)  \nonumber \\ &+\begin{pmatrix} 0 & 0 & 0 \\ 0 & 1 & 0 \\ 0 & 0 & 1  \end{pmatrix} G_{LR}(\omega,R)
\end{eqnarray}
where 
\begin{subequations}
\begin{align}
&G_{SR}(\omega,R) = \frac{e^{ikR}}{4\pi k^2 R^3}, \\
&G_{MR}(\omega,R) = \frac{i e^{ikR}}{4\pi k R^2}, \\
&G_{LR}(\omega,R) = \frac{i e^{ikR}}{4\pi R}
\end{align}
\end{subequations}
are the short, medium and long range contributions to the Green's dipole-dipole interaction, except for the polarization-dependent factor. As we see, the longitudinal modes don't have a long-range contribution. This is because dipoles don't radiate in the far-field in the direction of their axis. 

As we see in Equation (41) in the main text, we have, which in the 
\begin{align}
&\sum_{m\neq 0}\overleftrightarrow{\textbf{G}}(\omega,md)e^{-iqmd} = \nonumber \\ &\begin{pmatrix} 2 & 0 & 0 \\ 0 & -1 & 0 \\ 0 & 0 & -1 \end{pmatrix} \sum_{m\neq 0 } (G_{SR}(\omega,md)+G_{LR}(\omega,md)) e^{-iqmd}  \nonumber \\ &+ \begin{pmatrix} 0 & 0 & 0 \\ 0 & 1 & 0 \\ 0 & 0 & 1  \end{pmatrix} \sum_{m\neq 0}G_{LR}(\omega, md)e^{-iqmd}
\end{align}
where the lattice sums include interactions between all the particles. We evaluate these infinite lattice sums using polylogarithms $\mathrm{Li}_s(z)$ \cite{Pocock2018}, where $s$ is the order of the function and $z$ is a complex number. Polylogarithms are special functions usually implemented in scientific packages:
\begin{eqnarray}
\operatorname{Li}_s(z) = \sum_{m=1}^{\infty} \frac{z^m}{m^s}.
\label{eq:polylog}
\end{eqnarray}
where short, medium and long range terms are:
\begin{subequations}
\begin{align}
&\mathcal{G}_{SR}^{0}(\omega,q) = \sum_{m\neq 0}G_{SR}(\omega,q,md)e^{iqmd}=  \nonumber \\ &\frac{1}{4\pi k^2 d^3} \Big[\operatorname{Li}_3\left(e^{i(k-q)d}\right) +\operatorname{Li}_3\left(e^{i(k+q)d}\right)\Big] 
\\&\mathcal{G}_{MR}^{0}(\omega,q) = \sum_{m\neq 0}G_{MR}(\omega,q,md)e^{iqmd} = \nonumber \\&-\frac{k}{4\pi k d^2} \Big[\operatorname{Li}_2\left(e^{i(k-q)d}\right) +\operatorname{Li}_2\left(e^{i(k+q)d}\right)\Big] 
\\ &\mathcal{G}_{LR}^{0}(\omega,q) = \sum_{m\neq 0}G_{LR}(\omega,q,md)e^{iqmd}\nonumber \\ = &\frac{1}{4\pi d} \Big[\operatorname{Li}_1\left(e^{i(k-q)d}\right) +\operatorname{Li}_1\left(e^{i(k+q)d}\right)\Big].
\end{align}
\end{subequations}
\begin{align}
 &\sum\limits_{m}\overleftrightarrow{\textbf{G}}(\omega, md \pm \beta d/2) e^{iqmd}  =
 \nonumber \\ &\begin{pmatrix} 2 & 0 & 0 \\ 0 & -1 & 0 \\ 0 & 0 & -1 \end{pmatrix} \sum_{m} (G_{SR}(\omega,md \pm \beta d/2) \nonumber \\ &+G_{MR}(\omega,md\pm \beta d/2)) e^{iqmd}  + \nonumber \\ &\begin{pmatrix} 0 & 0 & 0 \\ 0 & 1 & 0 \\ 0 & 0 & 1  \end{pmatrix} \sum_{m}G_{LR}(\omega, md \pm \beta d/2 )e^{iqmd}
\end{align}
These sums can be evaluated using the Lerch trascendent, which is a generalization of polylogarithms and can be calculated as an integral: 
\begin{eqnarray}
\Phi(z.s,v) = \sum_{m=1}^{\infty} \frac{z^m}{(m+v)^s} =  \frac{1}{\Gamma(s)} \int_0^\infty \frac{t^{s-1} e^{-vt}}{1-ze^{-t}}\; dt,
\label{eq:Lerch_trascendent}
\end{eqnarray}

Then, the short, medium and long range sums are:
\begin{subequations}
\begin{align}
& \mathcal{G}^{\pm}_{SR}(\omega,q) =  \sum_{m} G_{SR}(\omega,q,  md \pm \beta d/2)e^{iqmd} = \nonumber \\ &
\frac{1}{4\pi k^2 d^3} \Big[e^{ik\beta d/2} \Phi\left( e^{i(k\pm q)d}, 3, \frac{\beta}{2}\right) \nonumber \\ &+ e^{i k \beta d/2}e^{i (k \mp q) a} \Phi\left( e^{i(k\mp q)d}, 3, 1 - \frac{\beta}{2}\right)\Big],  \\ & \mathcal{G}^{\pm}_{MR}(\omega,q) =  \sum_{m} (G_{MR}(\omega,q, md \pm \beta d/2 )e^{iqmd} = \nonumber \\  &-\frac{i}{4\pi k d^2} \Big[e^{ik\beta d/2} \Phi\left( e^{i(k\pm q)d}, 2, \frac{\beta}{2}\right)  \nonumber \\&+ e^{i k \beta d/2}e^{i (k \mp q) a} \Phi\left( e^{i(k\mp q)d}, 2, 1 - \frac{\beta}{2}\right)\Big],  \\
&\mathcal{G}^{\pm}_{LR}(\omega,q) = \sum_{m} (G_{LR}(\omega,q,md \pm \beta d/2)e^{iqmd} = \nonumber \\ & \frac{1 }{4\pi d} \Big[e^{ik\beta d/2} \Phi\left( e^{i(k\pm q)d}, 1, \frac{\beta}{2}\right) \nonumber \\ &+ e^{i k \beta d/2}e^{i (k \mp q) d} \Phi\left( e^{i(k\mp q)d}, 1, 1 - \frac{\beta}{2}\right)\Big].
\end{align}
\end{subequations}
The Green's matrix $\mathcal{G}(q)$ is now
\begin{align}
&\mathcal{G}(q) = \nonumber \\ &\begin{pmatrix} \mathcal{G}^{0}_{SR}(\omega,q) + \mathcal{G}^{0}_{MR}(\omega,q) &   \mathcal{G}^{+}_{SR}(\omega,q) + \mathcal{G}^{+}_{MR}(\omega,q) \\ \mathcal{G}^{-}_{SR}(\omega,q) + \mathcal{G}^{-}_{MR}(\omega,q) & \mathcal{G}^{0}_{SR}(\omega,q) + \mathcal{G}^{0}_{MR}(\omega,q) \end{pmatrix} \nonumber \\ &\otimes  \begin{pmatrix} 2 & 0 & 0 \\ 0 & -1 & 0 \\ 0 & 0 & -1 \end{pmatrix} \otimes \begin{pmatrix} 1 & 1 \\ 1 & 1 \end{pmatrix} \nonumber \\  &+ \begin{pmatrix} \mathcal{G}^{0}_{LR}(\omega,q) &   \mathcal{G}^{+}_{LR}(\omega,q)  \\ \mathcal{G}^{-}_{LR}(\omega,q)&  \mathcal{G}^{0}_{LR}(\omega,q) \end{pmatrix} \nonumber \\ &\otimes \begin{pmatrix} 0 & 0 & 0 \\ 0 & 1 & 0 \\ 0 & 0 & 1 \end{pmatrix} \otimes \begin{pmatrix} 1 & 1 \\ 1 & 1 \end{pmatrix} 
\end{align}
In the quasi-static limit, the topology of the bands depended exclusively on $\beta$. However, when we include all the interactions, we see the Green's matrix is not independent of the polarization and frequency anymore. References \cite{Pocock2018} and \cite{Pocock2019} showed that the long-range terms can produce topological transitions and the breaking of the bulk boundary correspondence. Consequently, beyond the quasi-static limit, the longitudinal modes will be less affected, which will make the edge states more robust compared to their transversal counterparts. 

Additionally, as the retardation and radiation corrections depend on $kd$, they will produce different effects in the modes depending on their frequency. The lower frequency bands will be expectedly less affected by these long-range corrections, which may make them more interesting in terms of robustness and topological protection. 

\bibliography{biblio_cs}
\end{document}